\documentclass[11pt,a4paper,logo,copyright]{googledeepmind}

\renewcommand{\today}{2026-09-11}

\AtBeginDocument{%
  }

\usepackage{multirow}
\usepackage{longfbox}
\usepackage{graphicx}
\usepackage{amsmath}
\usepackage{babel}
\usepackage{microtype}
\usepackage{subcaption}
\usepackage{dirtytalk}
\usepackage[ruled,linesnumbered]{algorithm2e}
\usepackage[most]{tcolorbox}
\usepackage{balance} 
\usepackage{enumitem} 
\usepackage{color}
\usepackage{url}
\usepackage{mwe}
\usepackage{wrapfig}
\usepackage{style/_custom}
\usepackage[authoryear,round]{natbib}
\providecommand{\Description}[2][]{}

\usepackage{xurl}
\usepackage{adjustbox}
\usepackage[nolist]{acronym}
\begin{acronym}
    \acro{HCI}{Human-Computer Interaction}
    \acro{CSCW}{Computer-Supported Collaborative Work}
    \acro{CS}{Computer Science}
    \acro{PCA}{Principal Component Analysis}
\end{acronym}

\newcommand{\human}[0]{Human-Only}
\newcommand{\ai}[0]{Standard Chatbot}
\newcommand{\unlock}[0]{Engage-to-Unlock}
\newcommand{\yoked}[0]{Time-Matched Unlock}

\newcommand{\tellme}[0]{\emph{Tell-me}}
\newcommand{\teachme}[0]{\emph{Teach-me}}

\begin{document}


\input{meta/authors}
\begin{abstract}
Generative AI can support writing, but frictionless access may cause cognitive offloading before users develop their own ideas.
We introduce \emph{\unlock}, a productive-friction mechanism that unlocks generative capabilities after users meaningfully engage with the task.
In a controlled experiment ($N = 398$), participants completed a writing task under one of four conditions--\human, \ai, \unlock, or \yoked, which matched unlock timing to \unlock\ participants but independent of users’ engagement--then evaluated passages for evidence and inferential errors.
Results show that \unlock\ redistributed effort across tasks: participants spent more time writing and less time evaluating, without increasing overall task duration.
They also submitted more prompts than in other AI-assisted conditions and showed the highest accuracy-per-time evaluation efficiency across conditions.
These findings suggest that designing GenAI access to encourage early human engagement may provide a productive form of friction, while retaining active AI use and efficient downstream evaluation.
\end{abstract}

\title{Who Thinks First? Designing Productive Friction with Engage-to-Unlock GenAI}

\makeatletter
\@ifclassloaded{acmart}{%

}{}
\makeatother


\maketitle
\section{Introduction}
Generative AI is transforming open-ended knowledge work by allowing users to delegate not only routine or surface-level tasks, but increasingly substantial cognitive work such as ideation, reasoning, problem formulation, and drafting~\cite{chi24-demands}. 
These capabilities can augment performance across a range of professional tasks: for example, access to ChatGPT has been shown to reduce completion time and improve output quality in professional writing tasks~\cite{science23-genai}.
Yet making such capabilities immediately and frictionlessly available also creates a risk of premature delegation.

Writing is not simply the production of text: writers develop positions, organize ideas, construct arguments, and determine how those ideas should be expressed~\cite{flower1981cognitive}. 
When generative assistance enters before this formulation has occurred, AI output can become the starting point for subsequent human work, potentially shaping ideas and directions users might otherwise develop themselves~\cite{chi25-timing}. Recent work characterizes this pattern as \emph{reactive writing}, in which writers evaluate and elaborate AI-proposed ideas before fully developing their own~\cite{chi26-reactive}.
AI assistants impact more than the workflow: opinionated language-model suggestions can shift both the views writers express and their subsequently reported attitudes \cite{chi23-opinionated}; AI suggestions can homogenize users’ writing toward Western cultural norms and styles \cite{chi25-homogenize}; and recent work finds that extensive LLM involvement can alter writers' intended meaning, stance, and perceptions such that the resulting text no longer reflects their own voice \cite{abdulhai2026llms}. 
These findings suggest that premature delegation may do more than reduce human effort: by introducing AI-generated ideas before writers have developed their own, it can shape the positions they form, the perspectives they express, and the extent to which the resulting text reflects their own intentions.

These concerns highlight the importance of having users develop their own understanding before encountering generative assistance. Prior HCI research has explored deliberately introduced friction---including cognitive forcing and delayed AI assistance---to preserve opportunities for independent engagement before AI input enters the task~\cite{cscw21-forcing, su-etal-2023-reviewriter, chi25-timing}. 
Such interventions can reduce overreliance and support more original ideation, but can also introduce usability and efficiency costs~\cite{cscw21-forcing, su-etal-2023-reviewriter}. 
These findings suggest that the benefit may lie not in delay itself, but in the opportunity it creates for users to develop an independent representation before engaging with AI-generated alternatives.

Friction may help users form an initial idea, position, or evaluative standard before encountering AI-generated alternatives. 
This independent representation provides a starting point for their work and a reference for evaluating AI output.
Yet friction is not inherently beneficial. 
Fixed delays may impose unnecessary costs on users who have already engaged with the task, while simply waiting does not ensure that meaningful engagement has occurred~\cite{cscw21-forcing, su-etal-2023-reviewriter}. 
This distinction motivates \emph{productive friction}: designing interaction constraints that support meaningful human engagement without unnecessarily obstructing AI assistance~\cite{natali2025better}. 
The design challenge, is not simply whether to add friction, but how to make that friction responsive to the human engagement it is intended to support.

We investigate \textbf{engagement-based GenAI access}: an interaction strategy in which generative capabilities become available based on evidence of task-relevant human contribution. 
We study this approach in argumentative writing, where early engagement can be reflected in the development of a position and supporting rationale.
Our implementation makes generative assistance available after writers have articulated these elements. 
We call this mechanism \textbf{\unlock}: a form of productive friction in which access to generative capabilities responds to each writer's engagement rather than elapsed time alone.

To investigate our mechanism, we conducted a controlled experiment with 398 participants across four access conditions.
In \human, participants completed the writing task without AI, providing a reference for fully human work.
In \ai, generative AI was available from the beginning, representing immediate, frictionless access. 
In both \unlock\ and \yoked, conversational AI assistance was available throughout the task, but direct text generation was initially restricted and unlocked later.
Under \unlock, unlocking responded to participants' task engagement. 
Under \yoked, unlock timing was matched to an \unlock\ participant but occurred independently of the current participant's engagement.

Participants first wrote an argumentative essay using four source summaries, then evaluated AI-generated passages grounded in the same sources without AI assistance. This second task complemented participants’ subjective appraisals by providing behavioral measures of how accurately and efficiently they evaluated AI-generated content. We ask two central research questions (RQs):
\begin{itemize}
    \item \textbf{RQ1.} Compared with \ai, what changes when \unlock~introduces engagement-based friction?
    \item \textbf{RQ2.} Compared with \yoked, does making that friction responsive to participants' own engagement add value beyond delay alone?
\end{itemize}

Our findings show that \unlock\ reshaped when and how participants invested effort. 
Compared with \ai, \unlock\ participants spent more time writing and reported greater required effort, while completing the subsequent evaluation faster.
Importantly, overall task duration did not differ significantly across conditions, suggesting that \unlock\ concentrated effort earlier in the workflow rather than simply extending it.
\unlock\ participants also interacted more extensively with AI, submitting more prompts than both \ai\ and \yoked.

The comparison with \yoked\ further suggests that these effects were not explained by delayed access alone. 
Despite receiving generative capabilities at comparable times, \unlock\ participants submitted more prompts and completed the subsequent evaluation faster, while maintaining comparable error-identification accuracy and showing stronger error-classification performance.
These differences identify where making friction responsive to users' engagement added value beyond time-based friction.

The key contributions are as follows:
\begin{itemize}
    \item We introduce \textbf{\unlock}, an interaction method that creates productive friction by initially restricting generative features and unlocking them in response to users' task engagement. 
    \item Our time-matched control separates friction that responds to engagement from friction caused by waiting, helping us distinguish the effects of adapting access to user engagement from the effects of delay alone. 
    \item We provide evidence that productive friction can change how people work with AI. \unlock\ encouraged users to engage more at the start without increasing total task time: they continued to use AI actively and evaluated the results efficiently. 
\end{itemize}

\section{Related Work}
\subsection{Generative AI, Cognitive Delegation, and Knowledge-Work Outcomes}

GenAI expands the scope of cognitive offloading in knowledge work, enabling users to delegate activities such as ideation, information synthesis, and reasoning~\cite{chi24-demands}.
Cognitive offloading uses external actions or resources to reduce internal processing demands, helping people overcome cognitive limitations and work more efficiently~\cite{risko2016cognitive}.
GenAI can improve productivity and output quality in professional writing~\cite{science23-genai} and yield particularly large productivity gains for less experienced workers~\cite{brynjolfsson2025generative}.
However, these benefits depend on task fit: \citet{dell2026navigating} found that AI-assisted knowledge workers were more likely to produce incorrect solutions on a task outside the system's capability frontier.

Beyond performance outcomes, GenAI can also redistribute rather than simply reduce human cognitive effort.
In a survey of knowledge workers, \citet{chi25-knowledge} found that greater confidence in GenAI was associated with lower self-reported critical-thinking effort, while workers described shifts from information gathering to verification, from problem solving to AI-response integration, and from task execution to oversight. 
In LLM-based search, users worked faster but were prone to over-relying on incorrect AI outputs~\cite{spatharioti2025effects}.
The tendency to delegate cognitive work to GenAI raises questions about how people exercise their own judgment and maintain reasoning and ideation skills in knowledge-work settings~\cite{chi25-knowledge}.
The concern, then, is not cognitive offloading itself, but when offloading displaces formative cognitive work: if users delegate ideation, reasoning, or judgment before developing their own understanding or position, they may lose opportunities for independent cognitive engagement.

Beyond productivity and correctness, research has examined how GenAI affects the creativity of individual outputs and the diversity of outputs across users.
These evaluations combine ratings of novelty and usefulness with computational measures of how similar or diverse the outputs are, often using distances between their semantic embeddings~\cite{doshi2024generative, anderson2024homogenization, ashkinaze2025ai}.
\citet{doshi2024generative} found that access to AI-generated story ideas improved evaluations of individual stories, particularly for writers with lower baseline creativity.
In argumentative writing, \citet{padmakumar2024does} found that co-writing with InstructGPT increased cross-author similarity and reduced lexical and key-point diversity.
Similarly, \citet{anderson2024homogenization} found that ChatGPT supported more numerous and elaborated ideas but reduced their semantic distinctiveness across users compared with a non-LLM creativity-support tool.
These effects are not uniform: in a dynamic ideation experiment, \citet{ashkinaze2025ai} found that high exposure to AI-generated examples increased collective semantic diversity without significantly changing individual creativity.
These findings motivate evaluating both individual artifact quality and cross-user content diversity when examining designs for GenAI access.
\subsection{Frictions in Human–AI Interaction}
While intelligent systems often aim to make assistance increasingly seamless, \ac{HCI} research has also explored deliberately introducing friction to preserve users' engagement in a task. 
Recent work on \emph{designed} or \emph{productive friction} instead considers deliberate interaction constraints that introduce effort or interruption in order to support engagement, reflection, critical reasoning, or user agency~\cite{chi26-design, natali2025better}. 
In human--AI interaction, such friction can take different forms.
Cognitive forcing functions require users to reason about a problem before viewing AI recommendations and can reduce overreliance on erroneous advice, although often with usability costs~\cite{cscw21-forcing}. Other interventions introduce reflective or metacognitive steps into AI-assisted decision making~\cite{cscw25-explanations, cscw23-explanations, ist25-metacognitive}. 
These works treat the conditions under which AI assistance becomes available or actionable as an interaction-design choice rather than assuming that more seamless access is always preferable.

Time itself can also serve as a form of friction. 
Slowing algorithmic responses can create opportunities for reflection~\cite{cscw19-slow}, while writing systems have delayed generated suggestions to preserve space for independent thought~\cite{su-etal-2023-reviewriter}. 
With generative AI, \citet{chi25-timing} found that writers who ideated independently before receiving LLM assistance produced more original ideas and reported greater creative self-efficacy than those with access from the outset. However, later assistance is not uniformly preferable. 
\citet{chi26-effects} found that the effects of LLM access timing depended on available task time: earlier access supported critical-thinking performance under time pressure but could impair it when participants had sufficient time. 
These findings underscore a central challenge for productive friction: constraints that create useful opportunities for independent engagement can also introduce unnecessary delay or interaction cost.

Existing interventions generally determine when such friction ends through a predefined criterion---for example, elapsed time, completion of a task stage, or a prescribed user action. 
If the value of delaying generative assistance lies partly in the engagement it makes possible, however, a fixed transition may be an imperfect proxy for whether that engagement has occurred. 
We therefore investigate a different approach: making access to generative capabilities based on observable task engagement. 
\unlock\ progressively releases generative capabilities as users engage with the task, rather than at a predetermined point in time. 
By comparing \unlock\ with \yoked, which introduces comparable temporal friction but unlocks independently of the current user's engagement, we examine whether engagement-based friction provides value beyond delayed access alone.
\section{Experiment}
We conducted a between-participants study with four conditions, recruiting participants through Prolific in two phases, to examine how access to generative AI affects argumentative writing and subsequent evaluation of AI-generated content. 
Participants first wrote an essay using shared source materials, then evaluated passages based on those sources. 
The passage-evaluation task complemented self-reported writing experience with a scored behavioral assessment, allowing us to examine performance separately from participants' appraisals of the assistance they received.
\human~provided an unaided writing baseline; \ai~made generative assistance available from the beginning; \unlock~made it available after contribution criteria were met; and \yoked~was designed to replay an \unlock~participant's unlock timing independently of the yoked writer's contribution. 
These comparisons examine what changes when generative access depends on prior engagement rather than being available immediately (\textbf{RQ1}), and whether this engagement-based access adds value beyond a matched delay (\textbf{RQ2}).

\subsection{Experimental Design and Conditions}
\begin{figure*}[t]
    \centering
    \includegraphics[width=\linewidth]{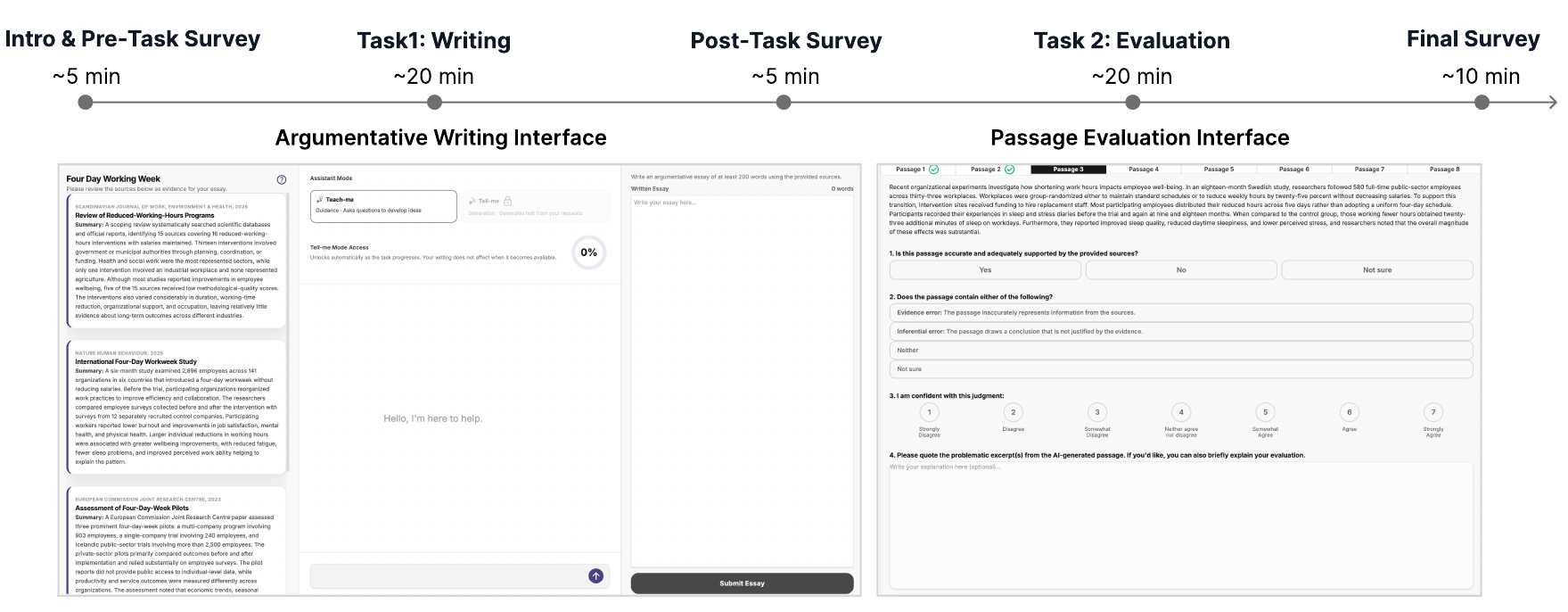}
    \caption{Study procedure and interfaces. 
    \textbf{Top:} Introduction and pre-task survey ($\sim$5 min), argumentative writing ($\sim$20 min), post-writing survey ($\sim$5 min), passage evaluation ($\sim$20 min), and final survey ($\sim$10 min). Task durations were suggested, not enforced. 
    Both tasks used the same four source summaries; passage evaluation was completed without AI assistance. 
    \textbf{Bottom left:} The \unlock~and \yoked~Access writing interface, with source materials, the conversational assistant and access indicators, and the essay editor. 
    The circular progress indicator reflected the writer's own draft development in \unlock~and replayed the matched \unlock~participant's progress on the same timeline in \yoked.
    \textbf{Bottom right:} The passage-evaluation interface, showing a passage, soundness judgment, error classification, 7-point judgment confidence, and an excerpt quotation field. The shared source-material panel remained available but is omitted from this screenshot.}
    \Description{A left-to-right timeline shows introduction and pre-task survey, Task 1 writing, post-task survey, Task 2 evaluation, and final survey. Below, the writing interface has three columns: shared source readings on the left, a conversational assistant with Teach-me and locked Tell-me modes and an unlock-progress indicator in the center, and an essay editor with a submit button on the right. The evaluation interface shows passage-navigation tabs, a passage, soundness choices, error-classification choices, a seven-point confidence scale, and a field for quoting a problematic excerpt. The two screenshots illustrate writing with an assistant followed by passage evaluation without AI assistance.}
    \label{fig:procedure}
\end{figure*}

\paragraph{Assistant Modes.}
This study used two types of AI assistants: the standard chatbot (which we call ``\tellme'') and a chatbot with a pedagogical prompt \citep{jurenka2024towards} triggering the Gemini ``Guided Learning'' mode \citep{team2024learnlm} (which we call ``\teachme''). 
Both types share the same \textit{gemini-3.1-pro-preview} model.
Both modes shared the task and conversational context; switching modes changed the response policy without starting a new conversation.
Appendix~\ref{sec:appendix_prompts} provides the complete system prompts, including the shared role and core instructions and the mode-specific policies for \teachme~and \tellme.

\paragraph{\unlock.}
\tellme~became available when the essay contained both (1) a position and (2) at least one distinct argumentative development linked to that position. 
A circular progress indicator displayed progress toward unlocking: a complete position and a supporting claim each contributed 50\%, while a partial or incomplete formulation of either contributed 25\%.
\tellme~became available when both elements were complete
and the indicator reached 100\%.
A banner announced availability of \tellme, but the interface remained in \teachme~until the participant explicitly selected \tellme. 
We detected the required elements using an LLM-based discourse classifier~\cite{calderon-etal-2025-alternative}.
We evaluated the classifier on 30 essays from the PERSUADE 2.0 corpus~\cite{24-persuade}, which provides discourse annotations for seven categories: \textit{Lead}, \textit{Position}, \textit{Claim}, \textit{Counterclaim}, \textit{Rebuttal}, \textit{Evidence}, and \textit{Concluding Statement}.
Against these reference annotations, the classifier achieved a micro-averaged $F_1$ of 91.5\% and a macro-averaged $F_1$ of 83.6\%.

\paragraph{\yoked.}
Prior evaluations of adaptive interfaces have used yoked controls to distinguish the effects of responding to a user's progress from those of delivering the same intervention on a matched schedule~\cite{dmello2016attending,mills2021eyemind}.
Following this approach, Phase~1 participants were randomly assigned to \human, \ai, or \unlock, and we recorded when \unlock~participants unlocked \tellme.
In Phase~2, we recruited a separate sample through Prolific, excluding anyone who had participated in Phase~1.
All Phase~2 participants were assigned to \yoked~and received an unlock schedule recorded from a Phase~1 \unlock~participant.
\yoked~participants began with \teachme, and \tellme~was scheduled to become available after the same elapsed writing time as for their matched participant, regardless of the \yoked~participant's own essay progress.
This comparison was intended to assess whether unlocking \tellme~based on the writer's own contribution offered benefits beyond delaying access for a matched duration. 

\paragraph{Condition Definitions.}
Participants in the \human~condition completed the task without access to the assistant. 
Participants in the \ai~condition had access to \tellme~throughout the writing task. 
Participants in the \unlock~and \yoked~conditions began with \teachme~then had access to \tellme~depending on their writing or progress of time.  

\subsection{Participants}
This study was reviewed and approved by our institute's ethics committee.
Participants were recruited via Prolific\footnote{\url{https://prolific.com}} across five English-speaking countries (Canada, India, South Africa, UK, USA) and received \$30 for the approximately one-hour study. 
This cross-national recruitment strategy provided a demographically diverse sample for evaluating the effects of AI access across user backgrounds.
Of 445 respondents who completed the session, 398 passed two embedded attention checks and were included in the final analysis (204 women, 184 men, and 10 participants in the combined non-binary, other, or unspecified category). 

Participants' mean age was 35.7 years ($SD = 11.5$, $Mdn = 33.0$, range 19--76); 40.7\% were aged 25--34, and 97.5\% held an undergraduate degree.
In Phase~1, participants were randomly assigned to \human, \ai, or \unlock. 
In Phase~2, all participants were assigned to \yoked. 
The final analysis included \human~($n = 99$), \ai~($n = 98$), \unlock~($n = 107$), and \yoked~($n = 94$).  

\subsection{Tasks and Materials}
\paragraph{Task 1: Argumentative Writing.}
Participants wrote at least 200 words about \textit{the four-day workweek}, drawing on summaries of four source documents. 
Two summaries presented supporting evidence and two presented critical evidence. 
We counterbalanced their order across four sequences that alternated evidence valence and study paradigm.

\paragraph{Task 2: Passage Evaluation.}
Task~2 assessed participants' ability to judge source-based content after the writing task.
Participants evaluated eight 150-word AI-generated passages based on the same four source summaries they had used in Task~1, with the source materials available for consultation and no AI assistance.
The task allowed us to assess both judgment accuracy and completion time: accuracy captured successful evaluation, while duration captured the time required to complete the assessment.
For each source, the item pool contained two sound passages and two flawed passages. 
Flaws represented two failure modes documented in prior model evaluations: \emph{evidence errors}, which misrepresented source evidence, and \emph{reasoning errors}, which drew conclusions not warranted by that evidence~\cite{jacovi-etal-2024-chain, kawabata-sugawara-2024-rationale}.
We instructed the AI to generate passages containing those errors, then manually validated the resulting passages.
Each participant evaluated four sound and four flawed passages, encountering one evidence-reporting item and one logical-reasoning item per source, with no passage repeated. 
Item soundness was counterbalanced across four experimental forms using a $4 \times 4$ Latin-square design.

\subsection{Procedure}
Figure~\ref{fig:procedure} summarizes the session. 
Participants first completed a survey on their English-language writing background and prior AI use. 
They then completed Task~1 under their assigned access condition, followed by a survey on writer workload, and the writing experience. 
During Task~1, clipboard transfers between the study interface and external windows were disabled in both directions to discourage participants from submitting task materials to external chatbots or importing externally generated text.
Participants could still copy text from the provided reading materials and, when available, the study assistant's responses into the essay editor.
Upon completing Task 1, they completed a post-task survey on writing agency and task load.
In Task~2, they evaluated the eight passages sequentially in their assigned counterbalanced order without the help of AI.
A final survey assessed the writing and evaluation experience.
Both tasks had a suggested duration of 20 minutes, but participants could finish at their own pace. 
Two attention checks~\cite{17-attention} appeared at the end of the pre-task and post-writing surveys.

\paragraph{Pilot Testing.}
Before the main study, six internal pilot sessions assessed instruction clarity, interface usability, and procedural flow. 
Two pilot participants also completed follow-up interviews. 
Their feedback informed revisions to the materials and procedure; pilot data were excluded from the main analysis.

\subsection{Measures and Analytic Strategy}
\label{sec:measures_analytic_strategy}
We assessed subjective experience, writing outcomes, passage evaluation performance, task duration and interaction behavior. 
The following subsections pair each measure with its scoring and analysis, using a shared inferential framework where applicable.

\subsubsection{Inferential Framework and Planned Comparisons}
\label{sec:inferential_framework}
Two planned comparisons corresponded directly to our research questions: \unlock~versus \ai~examined the effects of introducing engagement-based friction relative to \ai\ access (\textbf{RQ1}); \unlock~versus \yoked~examined whether making that friction responsive to participants' own engagement added value beyond delay alone (\textbf{RQ2}).

Unless otherwise specified, continuous participant-level outcomes were analyzed using Welch's one-way ANOVA and two-sided Welch's $t$-tests~\cite{delacre2017psychologists,04-note}.
Planned contrasts were evaluated regardless of omnibus significance. 
Across analyses, pairwise inference was restricted to these two comparisons, with Holm correction within each outcome ($m = 2$); component-specific and interaction families are specified below. 
We report effect estimates where available and distinguish adjusted ($p_{\mathrm{Holm}}$) from unadjusted ($p$) values.
Omnibus tests and descriptive summaries included all conditions. 
Omnibus tests were not adjusted across outcomes, and within-outcome corrections do not provide paper-wide family-wise error control. 
Exploratory analyses were identified separately. 

\subsubsection{Analytic Framework and Sample Inclusion}
Within the sample passing the attention checks, participants were analyzed according to their assigned condition, including \yoked~sessions affected by the unlock-delivery deviation reported in Results. 
Retaining these sessions avoided an additional exclusion based on intervention delivery that would selectively remove participants assigned longer unlock delays~\cite{fergusson2002post}.

\subsubsection{Writing Outcomes}
\label{sec:measures_writing}
\paragraph{Essay Length and Argumentative Structure.}
We measured final-essay word count and used the discourse classifier described above to derive the total number of discourse elements, category diversity (distinct categories present, 0--7), and element density per 100 words.
Omnibus discourse analyses used ordinary one-way ANOVAs across the four participant conditions ($N = 398$). The two planned pairwise contrasts compare \unlock~with \ai~and \yoked~using Welch's $t$-tests, with Holm correction across the two contrasts separately for each discourse outcome ($m = 2$). We report difference in means, standard errors, Welch degrees of freedom, and 95\% confidence intervals.

\paragraph{AI-Only Reference Corpus.}
We generated 100 independent essays using \textit{gemini-3.1-pro-preview}, approximately matching the sample size of one participant condition. 
Each essay was generated in an independent, single-turn context using the identical Task~1 instructions and the four source reading materials provided to human participants. 
To mirror the experimental interface and control for document presentation order, generations were evenly distributed across the four counterbalanced reading sequences (25 essays per order). 
Sampling parameters were held constant across all generations with default temperature $T = 1.0$\footnote{https://ai.google.dev/gemini-api/docs/gemini-3?hl=en}. 
This corpus serves as a descriptive reference representing zero human contribution.
It was not an additional randomized experimental condition and was excluded from the planned participant-condition contrasts. 
Exploratory \ac{PCA} comparisons also included this reference corpus; these comparisons characterize the observed corpora and do not estimate a randomized effect of AI-only authorship.

\paragraph{Semantic Content and Dispersion.}
After whitespace normalization, each complete final essay was embedded as a 3,072-dimensional vector using Gemini Embedding 2 with the \texttt{SEMANTIC\_SIMILARITY} task type. Within-condition semantic dispersion was the mean squared Euclidean distance from essay embeddings to their condition centroid:
\begin{equation}
    D_c = \frac{1}{n_c} \sum_{i=1}^{n_c} \|\mathbf{e}_i - \boldsymbol{\mu}_c\|^2,
    \qquad \boldsymbol{\mu}_c = \frac{1}{n_c} \sum_{i=1}^{n_c} \mathbf{e}_i,
\end{equation}
where $\mathbf{e}_i$ is an essay embedding and $n_c$ is the condition sample size. Lower dispersion indicates more concentrated semantic content. 
We report 95\% percentile bootstrap confidence intervals from 5,000 resamples and a homogenization index relative to \human:
\begin{equation}
    HI_c = 1 - \frac{D_c}{D_{\mathrm{\human}}}.
\end{equation}
Positive values indicate reduced dispersion relative to \human~writing; negative values indicate greater dispersion. These descriptive measures also included the AI-only reference corpus. Dispersion point estimates and bootstrap intervals were not used as a formal test of between-condition differences or equivalence.

To examine semantic location, embeddings from participant and AI-only essays (498 essays) were $L_2$-normalized, pooled, and analyzed using joint \ac{PCA} without condition labels. Component scores were compared exploratorily across all five groups using ordinary one-way ANOVAs. The two planned pairwise contrasts compare \unlock~with \ai~and \yoked~using Welch's $t$-tests, with Holm correction across the two contrasts within each tested component ($m = 2$), without correction across components. \ac{PCA} assessed location rather than diversity; exploratory $t$-SNE projections supported visualization only.

\paragraph{Lexical Frequency and Source Reuse.}
For lexical comparisons, essays were lowercased and tokenized into alphabetic words, excluding single-character tokens and closed-class stopwords. 
To account for variation in essay length across conditions, each token's raw count within an essay was normalized by the essay's total word count and scaled to occurrences per 1,000 words. 
We ranked the 85 most frequent content words in \human~and plotted each condition's mean normalized frequency (occurrences per 1,000 words; $\pm 1$ SEM across essays) in that shared order.

Because all conditions used the same reading materials, semantic overlap between \human~and AI-assisted essays could partly reflect direct reuse of shared source text. 
We conducted an exploratory source-reuse analysis to quantify the extent of verbatim borrowing across conditions.
Source texts and essays were stripped of HTML, lowercased, and tokenized.
Matches of at least eight consecutive words~\cite{rao2011external,sigmod03-winnowing} were extended while tokens matched; overlapping matches were merged to avoid double-counting. 
We report three measures: \textit{Explicit Paste}, the percentage of participants with at least one logged paste from the corresponding source into the essay editor, regardless of whether the pasted text remained in the submission; \textit{Verbatim Match}, the percentage of final essays containing at least one verbatim sequence match of eight or more consecutive words with the corresponding source; and \textit{Final Coverage}, the percentage of token positions in the final essay covered by a source match.
These measures distinguish paste actions during drafting, the prevalence of verbatim reuse across essays, and the extent of reuse within each final essay.

\subsubsection{Subjective Experience}
\label{sec:measures_subjective}
After Task~1, participants completed a survey assessing their writing experience and perceived workload to examine experiential trade-offs.
\paragraph{Agency and Ownership.}
Three single-item measures adapted from \citet{aied25-purposeful} assessed \textit{Process Control} (``I felt in control during the writing process''), \textit{Outcome Satisfaction} (``I feel content with the writing I produced''), and \textit{Document Ownership} (``I would feel comfortable publishing this essay under my name''). Participants rated each statement from 1 (\emph{Strongly disagree}) to 7 (\emph{Strongly agree}).

\paragraph{Workload and Disruption.}
Four items adapted from NASA-TLX~\cite{hart1988development} assessed mental demand, temporal demand, effort, and frustration, each rated from 1 (\emph{Very low}) to 7 (\emph{Very high}). 
Participants in the AI-assisted conditions also rated workflow disruption on a seven-point scale (``The way AI support was provided disrupted the flow of my writing'').

We analyzed each survey item separately using the continuous-outcome framework described above. 
Agency, satisfaction, ownership, and workload measures were compared across all four conditions; workflow disruption was compared across the three AI-assisted conditions.

\subsubsection{Passage Evaluation Performance}
\label{sec:measures_evaluation}
Task~2 yielded 3,184 trials ($398$ participants $\times$ $8$ passages). We distinguished two primary performance outcomes and one robustness outcome: (1) \textbf{Passage Evaluation Accuracy}: Across all trials, correctly accepting a sound passage or identifying a flawed passage was scored 1; incorrect and ``not sure'' responses were scored 0.
(2) \textbf{Error Classification Accuracy}: Among flawed passages correctly identified as flawed (779 trials), correct classification as an evidence or reasoning error was scored 1.
(3) \textbf{Complete Diagnosis Accuracy}: Across all flawed passages (1,592 trials), both detecting the flaw and correctly classifying its type was scored 1. This robustness outcome assessed whether conditional classification performance translated into successful diagnosis across all flawed items. The conditional outcome describes performance within the subset of detected flaws, whose composition can differ across conditions; it does not by itself establish an improvement across all flawed passages.

We fitted a binomial generalized linear mixed-effects model (GLMM) with a logit link for each outcome, using crossed random intercepts for participant and passage to account for repeated judgments:
\begin{equation}
\text{accuracy} \sim \text{condition} + (1 | \text{participant}) + (1 | \text{passage})
\end{equation}
Models were fitted using \texttt{glmer()} with \texttt{family = binomial(link = "logit")}. 
Here, \texttt{condition} was entered as a categorical fixed effect, while \texttt{participant} and \texttt{passage} were included as crossed random intercepts.
\unlock~was the reference condition, and models used the \textsc{bobyqa} optimizer. Contrasts used two-tailed Wald tests and the two-comparison Holm correction specified above. Results express contrasts as \unlock~minus the comparator, so positive coefficients and odds ratios greater than 1 favor \unlock; this reverses the sign of comparator coefficients from a model with \unlock~as the reference. We report contrast estimates, standard errors, odds ratios, 95\% family-wise Bonferroni-adjusted Wald confidence intervals, and Holm-adjusted $p$-values.

\subsubsection{Interaction Behavior}
\label{sec:measures_duration}
\paragraph{Task Duration and Allocation.}
We reconstructed Task~1 and Task~2 durations and their combined task time from logged events.
Because participants could work at their own pace, duration was treated as an outcome that could respond to the access manipulation.
For Task~2, duration indexed the time required to complete the eight-item assessment and was interpreted alongside judgment accuracy. Task~1 duration and combined task time provided context for whether faster evaluation was accompanied by greater time investment during writing.
In addition to the general condition comparisons, a repeated-measures linear mixed-effects model included fixed effects for condition, task, and their interaction, with a participant-level random intercept. 
Planned interaction contrasts compared the change in duration from writing to evaluation under \unlock~with the corresponding changes under \ai and \yoked. 
The two planned interaction contrasts form a separate Holm family ($m = 2$). These difference-in-differences characterized temporal trade-offs across the two tasks.
\human~durations are retained descriptively; the two reported pairwise comparisons are with \ai~and \yoked.

\paragraph{Interaction Logs.}
The writing interface recorded character insertions and deletions, copy/paste operations, and pause intervals. 
These logs supported reconstruction of revision trajectories, essay construction process and the integration of assistant suggestions.

\paragraph{Prompt Analysis}
To characterize how writers engaged with conversational AI, we analyzed $1{,}660$ user prompts across the three AI-assisted conditions ($n = 299$).
Interaction telemetry tracked chronological state transitions, attributing each query to either \teachme~mode or \tellme~mode. 

\section{Results}
\begin{figure*}[t]
    \centering
    \includegraphics[width=\linewidth]{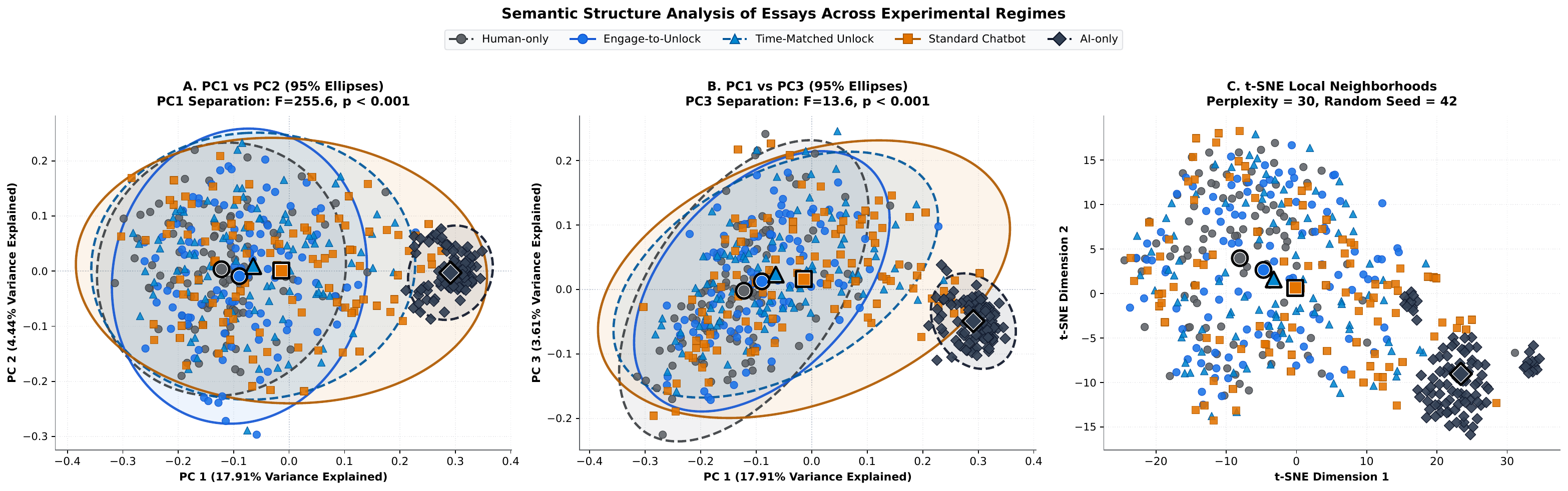}
    \caption{Global and local semantic representation of Task~1 essays for all conditions and the AI-only reference corpus, embedded via Gemini Embedding 2, $L_2$-normalized.
    Individual points denote individual essay submissions, shaded dashed contours represent 95\% bivariate normal confidence ellipses, and large black-outlined diamonds indicate condition centroids.
    \textbf{(A) PC1 vs.\ PC2 (Global Variation):} Principal Component 1 (17.91\% variance explained) shows separation across the five observed corpora in an exploratory, unadjusted omnibus test ($F(4, 493) = 255.62, p < .001$), with the AI-only centroid at $\mu = +0.291$ and participant-condition centroids at $\mu \in [-0.122, -0.014]$.
    PC2 (4.44\% variance explained) shows no significant condition separation ($F(4, 493) = 0.56, p = .693$).
    \textbf{(B) PC1 vs.\ PC3 (Secondary Structure):} Principal Component 3 (3.61\% variance explained; exploratory, unadjusted omnibus $F(4, 493) = 13.59, p < .001$) shows an additional difference across the five corpora; the reported AI-only centroid is $\mu = -0.049$.
    \ai~has a PC1 centroid ($\mu = -0.014$) between \human~($\mu = -0.122$) and the AI-only centroid.
    \textbf{(C) t-SNE Local Manifold Geometry:} Two-dimensional non-linear projection (perplexity $= 30$, seed $= 42$, initialized from the top 50 PCs). AI-only essays form a dense, isolated cluster in the lower-right quadrant ($10$-NN condition purity $= 98.0\%$). In contrast, all four participant conditions span a wide, overlapping manifold, with some \ai~essays positioned near the AI-only cluster in the projection.}
    \Description{Three scatterplots compare essays from \human, Engage-to-Unlock, Time-Matched Unlock, Standard Chatbot, and an AI-only reference corpus. Each point represents an essay; larger outlined symbols mark group centroids. Panels A and B plot principal component 1 against components 2 and 3, respectively, with group ellipses. Participant groups overlap broadly, while AI-only essays cluster at positive component-1 values. The Standard Chatbot centroid lies between the other participant centroids and the AI-only centroid on component 1. Panel C shows a t-SNE projection: participant essays occupy a broad overlapping region, while AI-only essays form a compact group toward the lower right, with some Standard Chatbot essays nearby.}
    \label{fig:pca_tsne}
\end{figure*}

\subsection{AI Availability and Use}
\subsubsection{Model Unlocking Status.}
In \unlock, 98 of 107 participants (91.6\%) unlocked \tellme, while 9 (8.4\%) did not reach the contribution criterion.
Of the 94 \yoked~participants, 60 received the unlock before submitting their essays. 
The remaining 34 did not receive an unlock before submission: 12 submitted their essays before their scheduled unlock time, and 8 were matched to \unlock~participants who never unlocked \tellme~and therefore remained in \teachme\ throughout the task. 
For the remaining 14 participants, the matched unlock was scheduled after 20 minutes, but a client-side timer unintentionally stopped replaying scheduled events at 20-minute. 
These participants therefore did not receive their scheduled unlock, although they could continue writing.

Among the 60 \yoked~participants who received an unlock, only 22 (36.7\%) had met the discourse criteria used by \unlock~when \tellme~became available. 
The remaining 38 (63.3\%) had not, including 16 participants whose drafts were still empty. 
Thus, although \yoked~reproduced the timing of unlocks from the \unlock~condition, it did not reproduce the engagement state that triggered those unlocks. 
In nearly two-thirds of cases, generative access became available before participants had articulated the position and supporting argument required by \unlock, illustrating the distinction between delaying access and making access dependent on meaningful task engagement.

\subsubsection{AI Interaction and Text Integration}
Participants in \unlock~submitted more prompts than those in \ai~and \yoked. Their mean final-essay coverage by verbatim AI-response text was lower than in \ai~and did not differ significantly from \yoked.

\paragraph{Participants in \unlock~Interacted More Frequently with the Assistant.}
We conducted the two planned Welch contrasts comparing \unlock~with \yoked~and \ai, with Holm correction within prompt count ($m = 2$).
Participants in \unlock~submitted more prompts ($M = 7.70$, $SD = 6.46$) than those in \yoked~($M = 5.40$, $SD = 5.12$), a difference of 2.30 prompts (Welch's $t(196.9) = 2.81$, $p_{\mathrm{Holm}} = .005$, $d = 0.39$).
\unlock~participants also submitted more prompts than those in \ai~($M = 3.35$, $SD = 2.98$), a difference of 4.35 prompts ($t(152.0) = 6.28$, $p_{\mathrm{Holm}} < .001$, $d = 0.85$).

\paragraph{Interaction Was Concentrated in \teachme~Even After \tellme~Became Available.}
\teachme~accounted for most prompts in both \unlock~(80.8\%) and \yoked~(76.2\%). 
Unlocking \tellme~did not necessarily lead to its use in either condition. 
Among participants who unlocked it, 46 of 98 in \unlock~(46.9\%) and 33 of 60 in \yoked~(55.0\%) subsequently submitted a \tellme~prompt. 
Twelve \unlock~participants and seven \yoked~participants returned to \teachme~after entering \tellme. 
Of these, seven \unlock~and five \yoked~participants returned after submitting at least one \tellme~prompt.

\paragraph{More Interaction Did Not Translate into More AI-Text Incorporation.}
At the condition level, \unlock~and \yoked~had higher mean prompt counts and lower mean verbatim AI-text coverage than \ai. These aggregate comparisons do not establish how prompt count relates to text incorporation within participants.
Participants submitted an average of 7.70 prompts under \unlock~and 5.40 under \yoked, compared with 3.35 under \ai, while final-essay coverage averaged 27.32\%, 24.51\%, and 47.16\%, respectively (Table~\ref{tab:overlap}). 
Planned Welch contrasts with Holm correction confirmed significantly lower AI text coverage under \unlock~(27.32\%) than \ai~(47.16\%; $p_{\mathrm{Holm}} = .002, d = -0.47$), while \unlock~and \yoked~did not significantly differ (24.51\%; $p_{\mathrm{Holm}} = .617, d = 0.07$).

\subsection{Content and Structure of the Essays}
We examined the final essays in terms of semantic content and diversity, source reuse and vocabulary, and the amount and structure of argumentation.
\subsubsection{Semantic Content and Diversity}
\paragraph{Semantic Dispersion.}
Table~\ref{tab:semantic_dispersion} reports within-condition semantic dispersion and homogenization relative to \human. Estimated dispersion ranged from 0.179 to 0.188 across the four participant conditions. 
\unlock~($D_c = 0.187$), \ai
($D_c = 0.188$), and \yoked~($D_c = 0.188$) each had slightly higher point estimates than \human~($D_c = 0.179$), corresponding to homogenization indices between $-0.050$ and $-0.045$.
The AI-only corpus had substantially lower estimated dispersion ($D_c = 0.065$), corresponding to 63.6\% lower dispersion than \human.

\begin{table}[t]
\centering
\caption{Semantic dispersion ($D_c$) and homogenization index ($HI_c$) across human writing conditions and the AI-only reference corpus.
Centroid dispersion ($D_c$) measures the mean squared Euclidean distance of essay embeddings to their condition centroid, with brackets reporting 95\% percentile bootstrap confidence intervals (5,000 resamples).
The Homogenization Index is defined relative to the \human~baseline ($HI_c = 1 - D_c / D_{\text{human}}$), where positive values reflect semantic compression.
The AI-only corpus has lower estimated dispersion ($HI_c = +63.6\%$); participant-condition estimates are descriptively similar. These intervals do not establish between-condition equivalence.}

\Description{Rows compare four participant conditions and an AI-only reference corpus. Columns report sample size, mean semantic dispersion with a 95 percent bootstrap confidence interval, and a homogenization index relative to Human-Only. Dispersion is 0.179 for Human-Only, 0.187 for Engage-to-Unlock, 0.188 for Time-Matched Unlock, 0.188 for Standard Chatbot, and 0.065 for AI-only. The three AI-assisted participant conditions have slightly negative homogenization indices, while AI-only has an index of positive 0.636, indicating 63.6 percent lower dispersion than Human-Only. Participant estimates are descriptively close; the intervals do not establish equivalence.}
\label{tab:semantic_dispersion}
\begin{tabular}{lccc}
\toprule
\textbf{Condition} & \textbf{$n$} & \textbf{\shortstack{Mean Semantic Dispersion ($D_c$)\\ \small [95\% Bootstrap CI]}} & \textbf{\shortstack{Homogenization Index\\ \small ($HI_c$ vs.\ \human)}} \\
\midrule
\human~        & 99  & 0.179 [0.167, 0.188] & \phantom{+}0.000 \small(baseline) \\
\unlock~   & 107 & 0.187 [0.178, 0.193] & -0.045 \small(-4.5\%)   \\
\yoked~     & 94  & 0.188 [0.177, 0.195] & -0.048 \small(-4.8\%)   \\
\ai~  & 98  & 0.188 [0.178, 0.194] & -0.050 \small(-5.0\%)   \\
\midrule
AI-only reference & 100 & 0.065 [0.061, 0.068] & \textbf{+0.636 \small(+63.6\%)}  \\
\bottomrule
\end{tabular}
\end{table}

\paragraph{Semantic Location.}
Whereas dispersion describes variation within each condition, semantic location describes where the essays are positioned relative to other conditions in the embedding representation.
Figure~\ref{fig:pca_tsne} shows the semantic locations of participant essays and AI-only essays.

Exploratory joint \ac{PCA} revealed that Principal Component 1 (PC1) explained 17.91\% of total semantic variance and separated the five observed corpora ($F(4, 493) = 255.62, p < 0.001$).
Among the participant conditions, \ai~had the PC1 centroid ($M = -0.014$) closest to that of the AI-only corpus ($M = +0.291$).
Essays composed under \unlock~($M = -0.090$) and \yoked~($M = -0.065$) had PC1 centroids descriptively closer to \human~($M = -0.122$) than \ai~did. 
Planned pairwise contrasts (Holm correction, $m = 2$ per component) showed that \unlock~differed significantly from \ai~along PC1 ($p_{\mathrm{Holm}} < .001$), whereas the contrast between \unlock~and \yoked~was not statistically significant ($p_{\mathrm{Holm}} = .104$). 
Contrasts along the remaining components did not detect significant differences between \unlock~and either comparison condition.
PC2 (4.44\% variance) showed no significant condition effect ($F = 0.56, p = 0.693$).
Complementary t-SNE visualization and descriptive $k$-nearest neighbor neighborhood purity showed a tightly clustered AI-only corpus, consistent with its lower estimated within-corpus dispersion.

\subsubsection{Source Reuse and Lexical Patterns.}
To help interpret these semantic patterns, we examined direct reuse of the shared readings and similarities in word usage.
\paragraph{Source Reuse.}
The semantic overlap between \human~and AI-assisted essays was accompanied by low levels of verbatim reuse of the shared readings.
Mean Final Coverage was 4.88\% in \human, 4.46\% in \unlock, 5.39\% in \yoked, and 2.94\% in \ai~(Table~\ref{tab:overlap}). 
Although 39.8\%--52.1\% of essays contained at least one source match, the low coverage values indicate that these matches generally occupied a limited portion of the text; 8.2\%--15.2\% of participants made an explicit paste from the readings. 
The AI-only reference combined a high match rate (74.0\%) with low mean coverage (2.03\%).

These findings provide little support for extensive verbatim copying of the shared readings as an explanation for the semantic overlap between \human~and AI-assisted essays in Figure~\ref{fig:pca_tsne}. 
This conclusion concerns direct text reuse: shared ideas or paraphrased source content could still contribute to the observed semantic overlap.

\begin{table*}[t]
\centering
\small
\caption{Verbatim text overlap metrics across all groups.
Length is the total essay words ($M \pm SD$).
\textbf{Source Document Usage} evaluates verbatim borrowing from source reading materials using an $n \ge 8$ sliding window.
\textbf{AI Response Usage} evaluates borrowing from chatbot responses via editor paste telemetry and sequence matching ($n = 299$).
\textit{Coverage} is the proportion of essay words matching the corresponding source ($M \pm SD$).
\textit{Match \%} is the percentage of essays containing at least one verbatim $n \ge 8$ sequence match.
\textit{Paste \%} is the percentage of participants who performed at least one logged clipboard paste from the source or chatbot into the essay editor, regardless of whether the pasted text remained in the final essay.
$^\dagger$Essays in the AI-only corpus were generated entirely by the model and have no interaction telemetry; \human~participants had no AI access (---).}
\Description{Rows compare Time-Matched Unlock, Engage-to-Unlock, Standard Chatbot, Human-Only, and an AI-only reference corpus. Columns report essay length and two groups of text-reuse measures: source-document usage and AI-response usage. Each group includes mean final-essay coverage, the percentage of essays with a verbatim match, and the percentage of participants with a logged paste. Mean source coverage ranges from 2.03 to 5.39 percent. Mean AI-response coverage is 24.51 percent for Time-Matched Unlock, 27.32 percent for Engage-to-Unlock, and 47.16 percent for Standard Chatbot. AI-only essays are longest, averaging 495.5 words. Dashes mark measures that are unavailable or not applicable.}
\label{tab:overlap}
\begin{adjustbox}{max width=\linewidth}
\begin{tabular}{l c ccc ccc}
\toprule
& & \multicolumn{3}{c}{\textbf{Source Document Usage}} & \multicolumn{3}{c}{\textbf{AI Response Usage}} \\
\cmidrule(lr){3-5} \cmidrule(lr){6-8}
\textbf{Condition} & \textbf{Length} & \textbf{Coverage} & \textbf{Match} & \textbf{Paste} & \textbf{Coverage} & \textbf{Match} & \textbf{Paste} \\
\midrule
\yoked~     & $285.6 \pm 96.7$  & 5.39\% $\pm$ 11.01\% & 52.1\% & 11.7\% & 24.51\% $\pm$ 39.70\% & 38.3\% & 25.5\% \\
\unlock~   & $299.2 \pm 101.3$ & 4.46\% $\pm$ 8.01\%  & 47.7\% & 12.1\% & 27.32\% $\pm$ 39.81\% & 46.7\% & 29.9\% \\
\ai~  & $308.3 \pm 138.9$ & 2.94\% $\pm$ 4.93\%  & 39.8\% & 8.2\%  & 47.16\% $\pm$ 43.84\% & 71.4\% & 40.8\% \\
\human~        & $277.1 \pm 102.7$ & 4.88\% $\pm$ 10.43\% & 46.5\% & 15.2\% & ---                   & ---    & ---    \\
AI-only reference & $495.5 \pm 38.3$  & 2.03\% $\pm$ 1.59\%  & 74.0\% & ---    & ---                   & ---    & ---    \\
\bottomrule
\end{tabular}
\end{adjustbox}
\end{table*}

\paragraph{Lexical Patterns.}
Figure~\ref{fig:token_frequency} compares mean normalized frequencies (occurrences per 1,000 words) for the 85 most frequent content words in \human. 
Several words used especially often in AI-only essays were also
used more often in AI-assisted essays than in \human~essays.
This descriptive pattern indicates similarities in word usage between AI-assisted essays and the AI-only reference. It does not distinguish the contribution of AI assistance from that of shared topics and source materials.

\subsubsection{Essay Length and Argumentative Structure}

Having examined semantic content and lexical patterns, we next considered whether access conditions also differed in essay length and argumentative structure.

\paragraph{Essay Length.} An analysis of essay length revealed that participants wrote an average of 277.11 words in the \human~ ($Mdn = 248.0$, $SD = 102.70$), 285.60 words in the \yoked~ ($Mdn = 239.5$, $SD = 96.67$), 299.21 words in the \unlock~($Mdn = 267.0$, $SD = 101.27$), and 308.32 words in the \ai~ ($Mdn = 253.5$, $SD = 138.92$). 
A Welch's one-way ANOVA indicated no statistically significant differences in essay word count across experimental conditions, $F(3, 216.55) = 1.43$, $p = .234$, with a negligible omnibus effect size ($\omega^2 = .004$, $\eta^2 = .012$). 
In contrast, the AI-only corpus contained substantially longer essays with lower variability ($Mdn = 496.0$, $SD = 38.34$), with nearly double the word count of the human-authored essays given the same instruction.

\subsubsection{Discourse Elements Diversity and Density.}
Exploratory ordinary one-way ANOVAs indicated a condition-level difference in total discourse elements ($F(3, 394) = 4.66$, $p = .003$). 
Mean counts were 8.30 for \human, 9.71 for \ai, 9.66 for \unlock, and 9.16 for \yoked. 
For counterclaims, \unlock~yielded significantly fewer counterclaims than \ai~($p_{\mathrm{Holm}} = .031$), but no significant difference from \yoked~($p_{\mathrm{Holm}} = .532$). 
For the remaining outcomes, contrasts between \unlock~and the two control conditions were not statistically significant.

\subsection{Passage Evaluation Performance}
\begin{figure}
    \centering
    \includegraphics[width=0.7\linewidth]{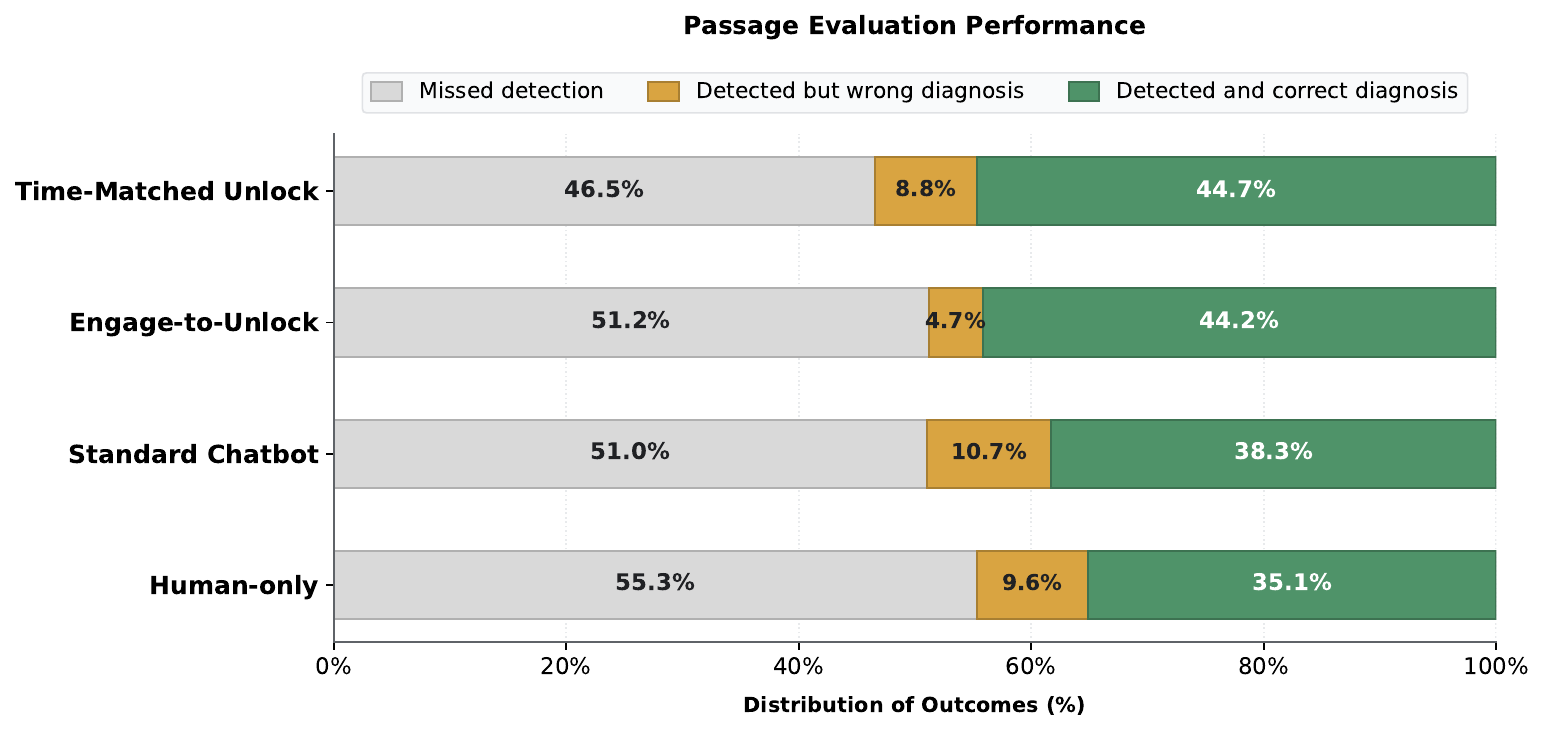}
    \caption{Distribution of flawed passage evaluation outcomes (1,592 trials across 398 participants), segmented into \emph{Missed detection} (gray), \emph{Detected but wrong diagnosis} (orange), and \emph{Detected and correct diagnosis} (teal). 
    While overall passage evaluation accuracy ranged across conditions ($61.6\%$--$66.4\%$), \textbf{\unlock} achieved markedly higher diagnostic accuracy once a flaw was identified ($90.4\%$ vs.\ $78.1\%$ in \ai; $p_{\mathrm{Holm}} = .003$).}
    \Description{Four horizontal stacked bars show percentages of flawed-passage trials with missed detection, detected but wrong diagnosis, and detected and correct diagnosis, respectively. Time-Matched Unlock: 46.5, 8.8, and 44.7 percent. Engage-to-Unlock: 51.2, 4.7, and 44.2 percent. Standard Chatbot: 51.0, 10.7, and 38.3 percent. Human-Only: 55.3, 9.6, and 35.1 percent. Gray, orange, and green segments encode these outcomes. Engage-to-Unlock has the smallest wrong-diagnosis segment, while its complete-diagnosis share is close to Time-Matched Unlock. Values are rounded and may not sum to exactly 100 percent.}
    \label{fig:task2}
\end{figure}

Task~2 assessed participants' judgments without AI assistance, providing behavioral outcomes alongside the self-reported writing experience. 
We report accuracy first, followed by completion time and efficiency.
We distinguish overall passage-evaluation accuracy, error-classification accuracy conditional on detecting a flaw, and complete-diagnosis accuracy across all flawed passages. 
Figure~\ref{fig:task2} summarizes the 1,592 flawed-passage trials, separating missed detections, detected flaws with incorrect classifications, and complete diagnoses.

\subsubsection{Judgment and Error-Classification Accuracy}

Participants detected similar numbers of flaws across conditions, with no significant differences between \unlock~and either comparator. Across all flawed passages, the proportion both detected and correctly classified was 44.16\% under \unlock, 38.27\% under \ai, and 44.68\% under \yoked; neither planned contrast was significant. 
Overall judgment accuracy likewise ranged from 61.61\% to 66.36\%, with no significant differences between conditions.

Among flaws that participants did correctly detect, however, classification accuracy differed. Participants in \unlock~classified detected errors more accurately than those in \ai~(90.43\% versus 78.12\%). 
The planned GLMM contrast indicated higher odds of correct classification under \unlock~($p_{\mathrm{Holm}} = .003$). 
Classification accuracy was also a bit higher than in \yoked~(83.58\%), although this contrast did not reach statistical significance ($p_{\mathrm{Holm}} = .051$). 
Thus, \unlock's advantage over \ai~was specific to classifying flaws once detected and did not translate into a statistically significant advantage in complete diagnosis across all flawed passages.

\subsubsection{Evaluation Time and Efficiency}
\paragraph{Faster Evaluation Under \unlock.}
\unlock~participants completed passage evaluation in 17.41 minutes on average, compared with 22.21 minutes under \ai~and 20.29 minutes under \yoked.
Planned contrasts showed shorter completion times relative to \ai~($\Delta M = -4.80$ min, $p_{\mathrm{Holm}} = .002$, $d = -0.48$) and \yoked~($\Delta M = -2.88$ min, $p_{\mathrm{Holm}} = .039$, $d = -0.30$).
\begin{figure*}[t]
    \centering
    \includegraphics[width=\linewidth]{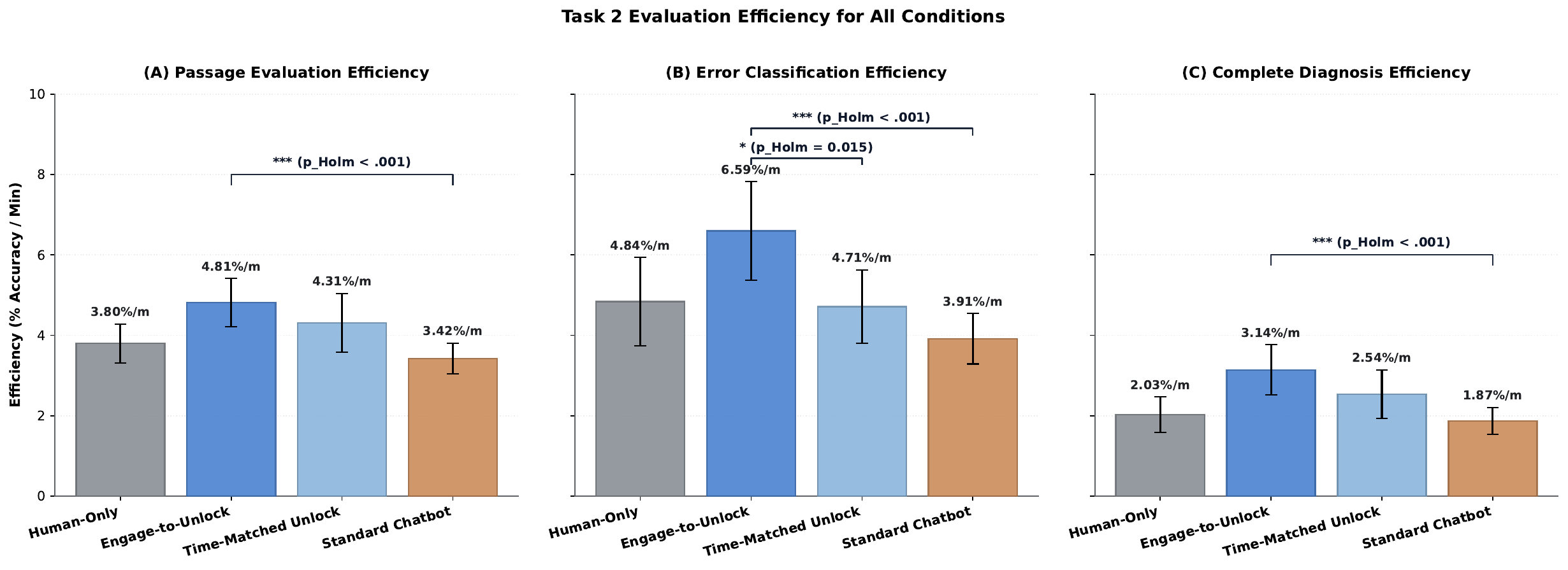}
    \caption{Task~2 evaluation efficiency across experimental conditions. Panels show condition means with 95\% confidence intervals for (A) passage evaluation efficiency, (B) conditional error classification efficiency, and (C) complete diagnosis efficiency. All panels report \% accuracy per minute of Task~2 duration. Panels A and C use fixed accuracy denominators of eight passages and four flawed passages, respectively, so their values are constant rescalings of correctly judged passages or complete diagnoses per minute. Panel B uses each participant's number of detected flaws as the accuracy denominator; it is not a constant rescaling of correctly classified flaws per minute. Upper-left boxes show omnibus Welch one-way ANOVAs across the four conditions. Brackets indicate planned pairwise contrasts comparing \unlock~with \ai~and \yoked~($^*p < .05$, $^{**}p < .01$, $^{***}p < .001$.}
    \Description{Three bar charts show mean accuracy percentage per minute with 95 percent confidence intervals. In Human-Only, Engage-to-Unlock, Time-Matched Unlock, and Standard Chatbot order, means are 3.80, 4.81, 4.31, and 3.42 for overall passage evaluation; 4.84, 6.59, 4.71, and 3.91 for error classification conditional on flaw detection; and 2.03, 3.14, 2.54, and 1.87 for complete diagnosis. Engage-to-Unlock has the highest mean in each panel. Significance brackets mark its advantage over Standard Chatbot in all three panels and over Time-Matched Unlock for conditional error-classification efficiency.}
    \label{fig:task2_efficiency}
\end{figure*}
\paragraph{Accuracy Relative to Evaluation Time.}
We next examined accuracy relative to evaluation time, measured as accuracy percentage per minute of Task~2 duration (Figure~\ref{fig:task2_efficiency}).
Compared with \ai, \unlock~showed higher values across all three measures: overall judgment efficiency ($p_{\mathrm{Holm}} = .0003$, $d = 0.53$), conditional error-classification efficiency ($p_{\mathrm{Holm}} = .0003$, $d = 0.60$), and complete diagnosis efficiency ($p_{\mathrm{Holm}} = .0010$, $d = 0.48$).
Relative to \yoked, the contrast was significant for conditional error-classification efficiency ($p_{\mathrm{Holm}} = .015$, $d = 0.38$), but not for overall judgment efficiency ($p_{\mathrm{Holm}} = .290$) or complete
diagnosis efficiency ($p_{\mathrm{Holm}} = .169$).

\subsection{Time Investment and Subjective Experience}
\begin{figure}
    \centering
    \includegraphics[width=0.5\linewidth]{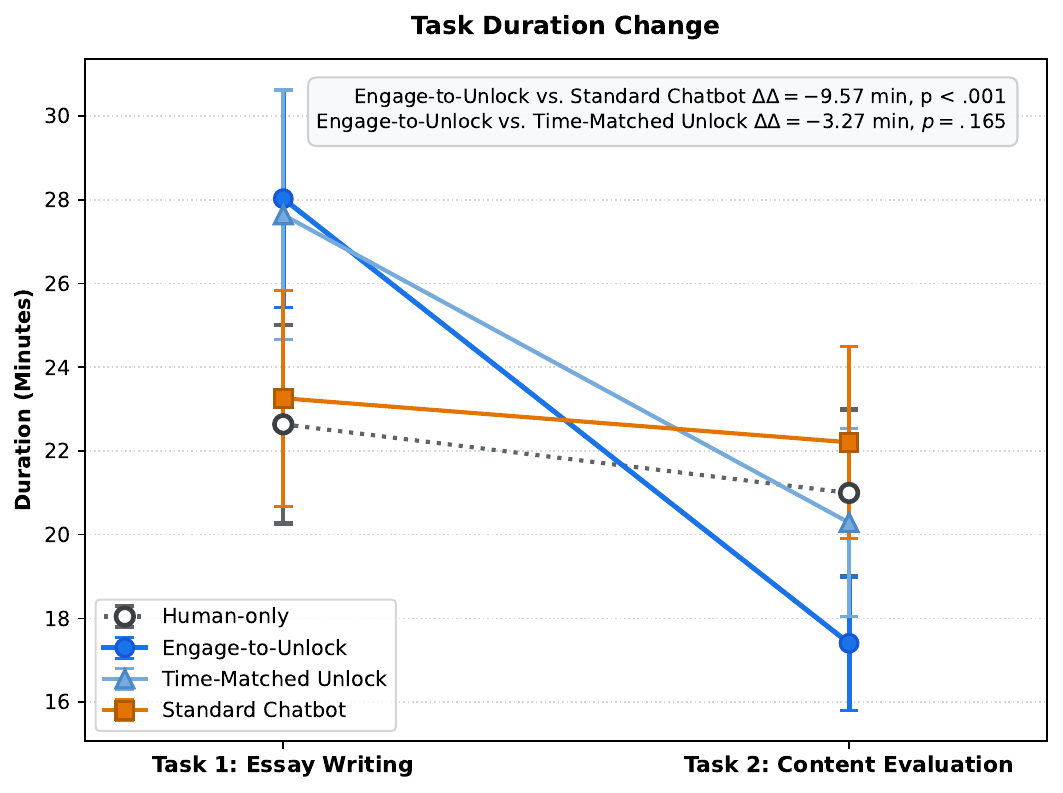}
    \caption{Mean task durations with 95\% confidence intervals across essay writing and passage evaluation for \human~($\circ$, dotted), \ai~($\blacksquare$), \yoked~($\blacktriangle$), and \unlock~($\bullet$). 
    A repeated-measures mixed model revealed a significant Condition $\times$ Task interaction ($F(3, 394) = 7.85, p < .001, \eta_p^2 = .056$). 
    In Task 2, \unlock~completed evaluations significantly faster than \ai~and \yoked~(Holm correction, $m = 2$).}
    \Description{A line chart connects mean writing and evaluation durations, with 95 percent confidence intervals. Engage-to-Unlock decreases from 28.02 to 17.41 minutes; Time-Matched Unlock from 27.64 to 20.29; Standard Chatbot from 23.26 to 22.21; and Human-Only from 22.63 to 20.99. Engage-to-Unlock starts with the longest mean writing time and ends with the shortest mean evaluation time. Its decrease is 9.57 minutes larger than Standard Chatbot's, a significant interaction contrast, and 3.27 minutes larger than Time-Matched Unlock's, a nonsignificant interaction contrast. Circles, triangles, squares, and distinct line styles distinguish the conditions.}
    \label{fig:task_duration_change}
\end{figure}

\paragraph{More Time Spent Writing, Less Time Spent Evaluating.}
\label{sec:time_allocation}
Total active task duration did not differ significantly across conditions ($F(3, 216.86) = 1.02$, $p = .383$).
Mean total duration was $43.63$ min ($SD = 14.77$) for \human, $45.46$ min ($SD = 16.51$) for \ai, $45.43$ min ($SD = 16.77$) for \unlock, and $47.92$ min ($SD = 18.91$) for \yoked.
However, time allocation differed when the two tasks were examined separately. 
For Task 1, duration differed significantly across conditions ($F(3, 217.30) = 4.69$, $p = .003$).
\unlock~($M = 28.02$ min, $SD = 13.52$) and \yoked~($M = 27.64$ min, $SD = 14.56$) showed longer mean durations than \ai~($M = 23.26$ min, $SD = 12.85$) and \human~($M = 22.63$ min, $SD = 11.88$). 
Durations between \ai~and \human~were close.
Planned pairwise comparisons using Welch’s $t$-tests with Holm correction ($m = 2$) confirmed that \unlock~took significantly longer than \ai~($+4.77\text{ min}$, $p_{\mathrm{Holm}} = .021$), but did not differ from \yoked~($+0.39\text{ min}$, $p_{\mathrm{Holm}} = .846$).

The pattern reversed in Task 2, where duration also varied significantly across conditions ($F(3, 214.31) = 4.87$, $p = .003$).
\unlock~had the shortest mean evaluation time ($M = 17.41$ min, $SD = 8.36$), compared with \yoked~($M = 20.29$ min, $SD = 10.94$), \human~($M = 20.99$ min, $SD = 9.97$), and \ai~($M = 22.21$ min, $SD = 11.46$). 
Planned pairwise comparisons (Holm-corrected, $m = 2$) revealed that \unlock~completed downstream evaluations significantly faster than both \ai~($-4.80\text{ min}$, $p_{\mathrm{Holm}} = .002$) and \yoked~($-2.88\text{ min}$, $p_{\mathrm{Holm}} = .039$).

As shown in Figure~\ref{fig:task_duration_change}, the change in duration from writing to passage evaluation differed across conditions. 
The Condition $\times$ Task interaction was significant, $F(3,394)=7.85$, $p<.001$, indicating that the change in duration from Task 1 to Task 2 differed across conditions.

Planned interaction contrasts, Holm-adjusted as a two-comparison family ($m = 2$), showed a pronounced difference between \unlock~and \ai.
Adaptive participants spent 4.77 min longer on Task 1 but 4.80 min less on Task 2, yielding a significant cross-task difference-in-differences of $-9.57$ min ($p_{\mathrm{Holm}}<.001$).
In contrast, \unlock~and \yoked~showed a smaller difference-in-differences of $-3.27$ min, which was not significant ($p_{\mathrm{Holm}}=.165$).
Thus, \unlock~substantially redistributed time across tasks relative to \ai, whereas its temporal allocation did not significantly differ from \yoked.
\human~durations provide descriptive context.

\paragraph{Participants in \unlock~Reported Greater Effort Than Those in \ai.}
To evaluate task workload, we conducted a Welch’s one-way ANOVA across the four experimental conditions. 
The omnibus test revealed a statistically significant difference in required effort ($F(3, 215.85) = 4.72, p = 0.003, \eta^2 = 0.040, \omega^2 = 0.032$). 
In contrast, no significant omnibus differences emerged for mental demand ($F(3, 217.61) = 1.35, p = 0.259$), temporal demand ($F(3, 218.52) = 0.57, p = 0.633$), or frustration ($F(3, 218.11) = 0.95, p = 0.419$). 

Planned contrasts with Holm correction showed that participants in \unlock~reported significantly greater effort than those in \ai~($M = 5.15 \pm 1.14$ vs. $M = 4.47 \pm 1.48$; $p_{\mathrm{Holm}} < 0.001$).
However, no significant difference in reported effort was detected between \unlock~and \yoked~($p_{\mathrm{Holm}} = 0.585$).
Participants in \unlock~reported greater effort than those in \ai, while no statistically significant differences between these conditions were detected in frustration ($p_{\mathrm{Holm}} = .358$) or temporal demand ($p_{\mathrm{Holm}} = 1.000$).

\paragraph{No Significant Differences Detected in Agency, Satisfaction, Ownership, or Disruption.}
The two planned contrasts comparing \unlock~with \ai~and \yoked~revealed no significant differences in process control, outcome satisfaction, or document ownership after Holm adjustment.
For process control, the contrast between \unlock~and \ai~ was positive but not statistically significant ($p_{\mathrm{Holm}} = .085$).

A Welch's one-way ANOVA detected no significant difference in perceived workflow disruption across the three AI-assisted conditions ($F(2, 194.35) = 0.056$, $p = .946$).
Mean ratings on the seven-point scale were close: \unlock~($M = 3.17$, $SD = 1.66$), \ai~($M = 3.09$, $SD = 1.75$), and \yoked~($M = 3.12$, $SD = 1.81$).
We found no evidence that \unlock~disrupted participants' writing flow more than the other AI access conditions.

\section{Discussion}

\subsection{Designing Where Humans Think: From Effort Reduction to Effort Reallocation}
Prior work has argued that generative AI can shift human cognitive work from producing content toward evaluating, verifying, and integrating model outputs \cite{chi25-knowledge, chi24-demands, sarkar2023exploring}. 
However, our findings suggest that this downstream evaluative work should not necessarily be interpreted as productive effort. 
Evaluating AI output itself depends on having sufficient task and domain knowledge to judge what the model has produced \cite{chi24-demands}. 

Viewed through this lens, the additional evaluation time observed under \ai\ may reflect, in part, cognitive work that was deferred rather than eliminated. 
Because participants could delegate generation immediately, they may have developed less task-specific understanding during the initial writing task and therefore needed more time to establish a basis for judging AI-generated content later. 
By contrast, \unlock\ required participants to engage with the task before generative assistance became available, potentially front-loading some of the sensemaking needed for subsequent evaluation.
Consistent with this interpretation, \unlock\ participants completed the evaluation task faster without a corresponding reduction in evaluation performance.

This possibility suggests a different way of thinking about where friction should be introduced in human--AI workflows.
Prior work address AI overreliance primarily through cognitive forcing functions, which introduce friction at the point of AI-assisted judgment \cite{cscw21-forcing}. 
Such interventions effectively ask users to expend \emph{more} effort when scrutinizing AI. 
\unlock~instead suggests a complementary strategy: rather than uniformly increasing friction, systems may redistribute engagement toward stages where independent human contribution is particularly valuable. 
The objective is therefore neither maximal automation nor maximal human effort, but a more deliberate allocation of effort across the workflow.

More broadly, this motivates a shift from designing AI systems primarily for \emph{effort reduction} toward designing for \emph{effort allocation}~\cite{memmert2025effort}. 
Human--AI interfaces might explicitly consider which stages benefit most from independent human reasoning, authorship, or judgment, and selectively preserve engagement there while allowing AI to reduce effort elsewhere. In this sense, an important design problem for human--AI interaction is not only determining \emph{how much} humans should think, but \emph{where} they should think.


\subsection{Beyond Binary AI Access: Capability Boundaries in Human–AI Collaboration}
Generative AI can improve what people accomplish with assistance while also replacing activities that may matter for later independent performance~\cite{liu2026ai}. 
In educational settings, unrestricted chatbot access can improve performance while assistance is available yet impair subsequent independent performance, while carefully designed safeguards can preserve benefit while mitigating negative effects \cite{pns25-genai}.
Similar tensions appear in writing: ChatGPT assistance can improve produced essays without corresponding gains in knowledge or transfer \cite{bet25-laziness}.
At the same time, these effects are not uniformly erosive. Other experiments show that AI-supported writing can improve later unassisted performance when AI provides useful examples from which users can learn \cite{lira2025coach}.
These findings suggest that whether AI acts as an \emph{enhancer} or \emph{eroder} depends not simply on whether it is present, but on the way it provides assistance.
We build on this distinction by considering a more specific design question: whether access should be defined by the AI system as a whole, or by the particular capabilities it provides.

\unlock~demonstrates one way of separating these notions. 
Rather than withholding AI altogether or changing the underlying model, it conditions access to a particular capability---direct generative production---while preserving opportunities for AI-supported interaction. 
Our findings suggest that such a boundary can alter where participants invest their effort, with greater reported effort than \ai~but no detected increase in overall task time.
This capability view builds on longstanding work showing that automation need not be treated as all-or-none: different functions can be automated to different degrees, with different consequences for human activity \cite{parasuraman2000model}.

For generative AI, this opens a broader design space in which different capabilities need not become available simultaneously. For example, early in a task, a system might support retrieval, questioning, or reflection without directly producing the target output; after users have formed an initial representation, it might support critique, comparison, or explanation; later, direct generation could support rewriting, synthesis, or automation.
Capability boundaries also need not be fixed solely by task stage. 
Future systems could adapt them based on \emph{demonstrated engagement}, uncertainty, expertise, task stakes, learning objectives, or prior patterns of reliance. 
A novice learning a skill, for example, may benefit from different boundaries than an expert optimizing production; similarly, a system designed for learning may appropriately expose different capabilities than one designed purely for efficiency. 
This framing shifts the design question from \say{Should this user have access to AI?} toward \say{What should this AI be able to do for this user, at this moment?} 
And these mechanisms should be understood as scaffolding rather than prohibition.
Future work could therefore evaluate capability boundaries across both longer timescales and larger populations: not only whether they preserve immediate performance, but whether they affect learning, transfer, independent judgment, authorship, and collective diversity. 

\subsection{Limitations and Future Work}
A timer constraint affected unlock delivery for 14 of 94 \yoked~participants, creating a partial mismatch in generative access between the \unlock~and \yoked~conditions. 
This delivery deviation limits how precisely the \unlock~vs. \yoked~comparison isolates the effect of contribution-based access. In addition, \yoked~participants were recruited in a separate second phase, whereas Phase~1 participants were randomly assigned to \human, \ai, or \unlock. Differences involving \yoked~may therefore also reflect recruitment-phase or cohort differences; the comparisons among the three Phase~1 conditions retain random assignment.
Our findings also come from a short-term, writing-focused experiment. 
Although \unlock~changed engagement and downstream evaluation, it remains unclear whether these effects persist, transfer to later unassisted performance, or generalize beyond writing. 
Longitudinal studies should therefore examine effects on learning, skill retention, and transfer across repeated use and other domains.
Finally, \unlock~represents only one form of capability boundary. 
Future systems could vary which capabilities are exposed, when they become available, and whether access depends on task stage, expertise, uncertainty, or demonstrated engagement.

\section{Conclusion}
We investigated contribution-based access to generative AI in an argumentative writing experiment with 398 participants. 
\unlock~made generative assistance available after writers expressed a position and supporting argument. 
Compared with \ai~Access, participants spent more time writing, reported greater effort, and completed subsequent passage evaluation faster, with no significant differences in frustration or temporal demand. 
Among correctly detected flaws, \unlock~participants classified errors more accurately than those in \ai, while the comparison with \yoked~was not statistically significant.
\unlock~participants also submitted more prompts than those in \ai~and \yoked, mainly in \teachme~mode, potentially indicating greater engagement with conversational guidance. 
Fewer than half of \unlock~participants who gained access to \tellme~subsequently prompted it.
These findings motivate designing AI access around users' contributions and examining how capability boundaries shape engagement and effort across tasks. 
Future work could investigate their longer-term effects on learning, independent judgment, and skill retention.
\section*{Acknowledgements}
This work was completed during my student researcher internship at Google DeepMind in London, an experience made unforgettable by a community of incredibly passionate and inspiring researchers. 
I deeply enjoyed collaborating with my research group and want to thank everyone for the rich learning environment and shared insights.

My deepest gratitude goes to Piotr Mirowski for his mentorship, constant support, and guidance from start to finish. 
I am also indebted to Tal Linzen and Hannah Rashkin, whose contributions to a prior project served as the spark for this research. 
I highly appreciate Shakir Mohamed for his attentive, kind, and steady support. 
Lastly, my thanks to Katie Stasaski for her excellent prototype feedback and continuous support, and to Irina Jurenka for her assistance in bringing this paper to completion.

\bibliographystyle{ACM-Reference-Format}
\bibliography{ref}
\newpage
\appendix
\section{Demographics Information}
\begin{table*}[t]
\centering
\small
\caption{\textbf{Participant Demographic Characteristics across Experimental Conditions.} Values are counts and within-condition percentages unless otherwise noted ($N = 398$). No statistically significant differences in demographic characteristics were detected across conditions (all $p > .30$; one-way ANOVA for age and Pearson $\chi^2$ tests for categorical measures).}
\Description{Participant characteristics are organized by age, gender or sex, country, employment status, and education. Columns show Engage-to-Unlock, Standard Chatbot, Human-Only, Time-Matched Unlock, and the overall sample, corresponding to 107, 98, 99, 94, and 398 participants. Age is reported as mean and standard deviation; categorical entries show counts and percentages. Overall mean age is 35.7 years. The sample includes 204 female participants, 184 male participants, and 10 in the combined non-binary, other, or unspecified category. Participants come from Canada, India, South Africa, the United Kingdom, and the United States. Half report full-time employment, and 97.5 percent are listed in the undergraduate-degree category.}
\label{tab:full_demographics}
\begin{adjustbox}{max width=\linewidth}
\begin{tabular}{lccccc}
\toprule
\textbf{Demographic Characteristic} & \textbf{\unlock} & \textbf{\ai} & \textbf{\human} & \textbf{\yoked} & \textbf{Total} \\
 & $(107)$ & $(98)$ & $(99)$ & $(94)$ & $(398)$ \\
\midrule
\textbf{Sample Size} & 107 (26.9\%) & 98 (24.6\%) & 99 (24.9\%) & 94 (23.6\%) & 398 (100.0\%) \\
\addlinespace
\textbf{Age (years)}, Mean $\pm$ SD & 36.0 $\pm$ 10.2 & 36.9 $\pm$ 12.9 & 35.7 $\pm$ 11.8 & 34.0 $\pm$ 11.1 & 35.7 $\pm$ 11.5 \\
\addlinespace
\textbf{Gender / Sex, $n$ (\%)} & & & & & \\
\quad Female & 61 (57.0\%) & 49 (50.0\%) & 46 (46.5\%) & 48 (51.1\%) & 204 (51.3\%) \\
\quad Male & 42 (39.3\%) & 46 (46.9\%) & 51 (51.5\%) & 45 (47.9\%) & 184 (46.2\%) \\
\quad Non-binary / Other / Unspecified & 4 (3.7\%) & 3 (3.1\%) & 2 (2.0\%) & 1 (1.1\%) & 10 (2.5\%) \\
\addlinespace
\textbf{Geographic Cohort / Country, $n$ (\%)} & & & & & \\
\quad Canada (CA) & 21 (19.6\%) & 17 (17.3\%) & 17 (17.2\%) & 16 (17.0\%) & 71 (17.8\%) \\
\quad India (IN) & 8 (7.5\%) & 8 (8.2\%) & 8 (8.1\%) & 13 (13.8\%) & 37 (9.3\%) \\
\quad South Africa (SA) & 15 (14.0\%) & 15 (15.3\%) & 13 (13.1\%) & 14 (14.9\%) & 57 (14.3\%) \\
\quad United Kingdom (UK) & 18 (16.8\%) & 20 (20.4\%) & 18 (18.2\%) & 18 (19.1\%) & 74 (18.6\%) \\
\quad United States (US) & 45 (42.1\%) & 38 (38.8\%) & 43 (43.4\%) & 33 (35.1\%) & 159 (39.9\%) \\
\addlinespace
\textbf{Employment Status, $n$ (\%)} & & & & & \\
\quad Full-Time & 54 (50.5\%) & 47 (48.0\%) & 50 (50.5\%) & 49 (52.1\%) & 200 (50.3\%) \\
\quad Part-Time & 16 (15.0\%) & 16 (16.3\%) & 18 (18.2\%) & 21 (22.3\%) & 71 (17.8\%) \\
\quad Unemployed (job seeking) & 13 (12.1\%) & 9 (9.2\%) & 17 (17.2\%) & 11 (11.7\%) & 50 (12.6\%) \\
\quad Not in paid work & 10 (9.3\%) & 5 (5.1\%) & 3 (3.0\%) & 4 (4.3\%) & 22 (5.5\%) \\
\quad Other / Job starting soon & 6 (5.6\%) & 9 (9.2\%) & 6 (6.1\%) & 4 (4.3\%) & 25 (6.3\%) \\
\quad Unknown / Unspecified & 8 (7.5\%) & 12 (12.2\%) & 5 (5.1\%) & 5 (5.3\%) & 30 (7.5\%) \\
\addlinespace
\textbf{Education Level, $n$ (\%)} & & & & & \\
\quad Undergraduate degree (BA/BSc/other) & 104 (97.2\%) & 95 (96.9\%) & 97 (98.0\%) & 92 (97.9\%) & 388 (97.5\%) \\
\quad Graduate degree (MA/MSc/PhD) & 0 (0.0\%) & 0 (0.0\%) & 0 (0.0\%) & 1 (1.1\%) & 1 (0.3\%) \\
\quad Unknown / Unspecified & 3 (2.8\%) & 3 (3.1\%) & 2 (2.0\%) & 1 (1.1\%) & 9 (2.3\%) \\
\bottomrule
\end{tabular}
\end{adjustbox}
\end{table*}

\section{Task Duration Distributions}
\begin{figure*}[t]
    \centering
    \includegraphics[width=\linewidth]{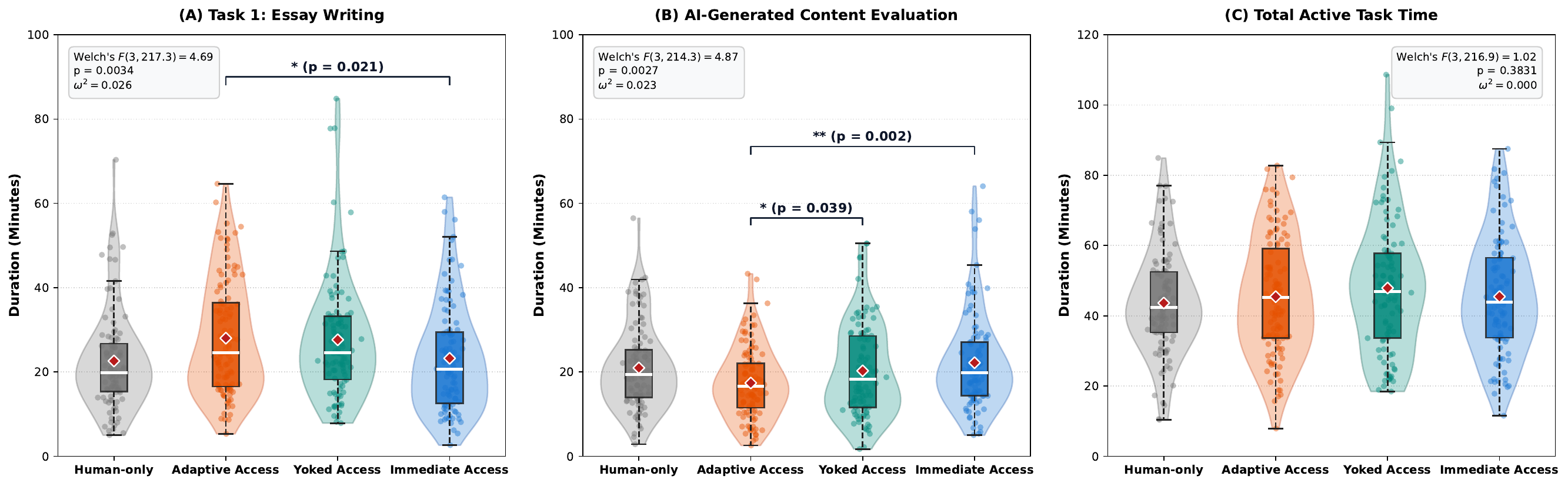}
    \caption{Task-duration distributions for (A) essay writing, (B) passage evaluation, and (C) combined task time. 
    Violins show kernel densities, dots show individual participants, boxplots show medians and interquartile ranges, and red diamonds show means. Significant pairwise annotations in panels A and B report Welch tests comparing \unlock~with \ai~and \yoked, with Holm correction across two comparisons per task. No significant condition difference was detected in combined task time.}
    \Description{Three panels compare task durations across Human-Only, Engage-to-Unlock, Time-Matched Unlock, and Standard Chatbot. Violin shapes show distributions, dots represent participants, boxplots mark medians and interquartile ranges, and red diamonds mark means. For essay writing, Engage-to-Unlock and Time-Matched Unlock have longer mean durations than Human-Only and Standard Chatbot. For passage evaluation, Engage-to-Unlock has the shortest mean duration. Pairwise annotations indicate that Engage-to-Unlock takes significantly longer than Standard Chatbot for writing but significantly less time than both Standard Chatbot and Time-Matched Unlock for evaluation. No significant condition difference is detected for combined task time.}
    \label{fig:task_duration_distributions}
\end{figure*}

Figure~\ref{fig:task_duration_distributions} shows task-duration distributions across conditions.

\section{Lexical Frequency Patterns}
\begin{figure*}[t]
    \centering
    \includegraphics[width=\linewidth]{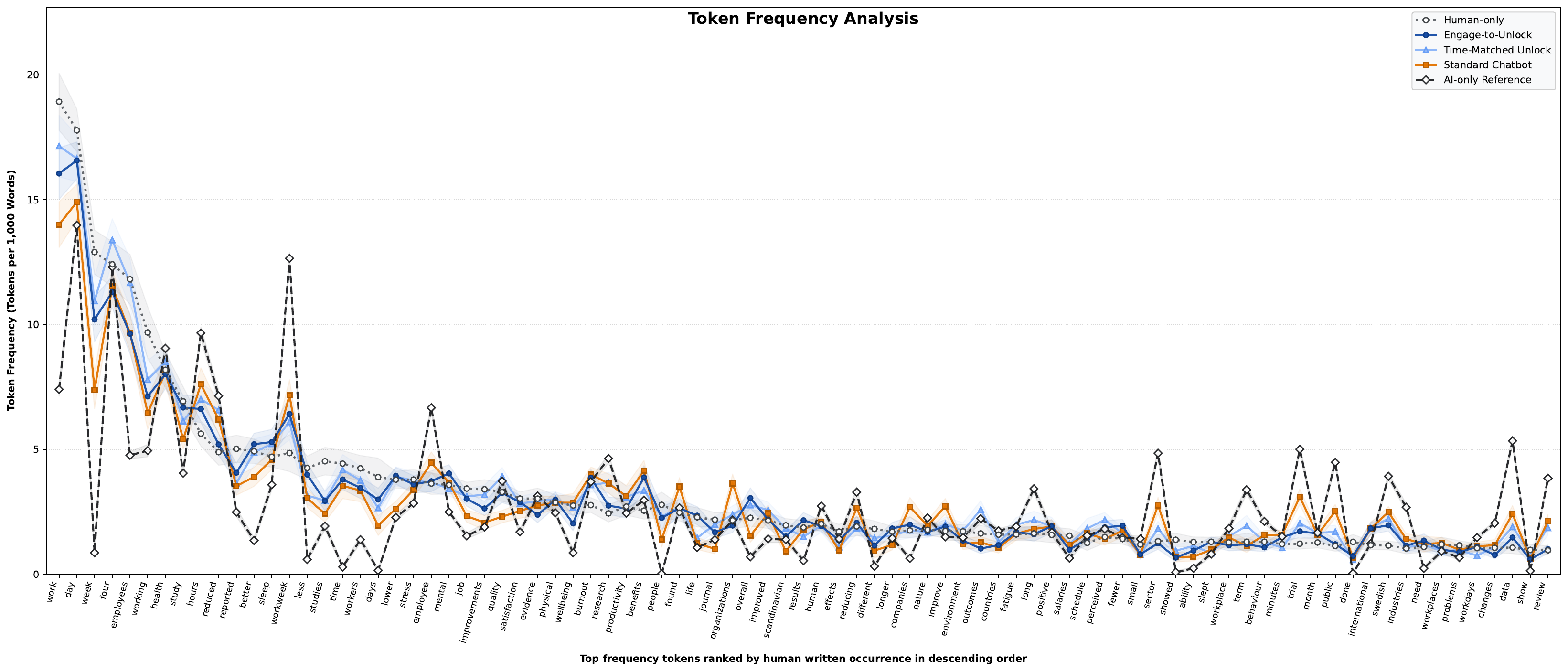}
    \caption{Mean normalized frequencies (occurrences per 1,000 words) of the 85 most frequent content words in \human, ordered by descending \human~frequency. 
    Points show condition means across essays and shaded bands indicate $\pm 1$ SEM. 
    Single-character tokens and closed-class stopwords were excluded. Each token's count was divided by the essay's total word count and multiplied by 1,000 before averaging across essays.}
    \Description{A line chart compares five corpora across 85 content words, ordered from most to least frequent in Human-Only essays. The horizontal axis lists words; the vertical axis measures occurrences per 1,000 words. Points show condition means and shaded bands show one standard error on either side. Human-Only, Engage-to-Unlock, Time-Matched Unlock, and Standard Chatbot generally follow similar declining profiles. The AI-only reference varies more sharply across words, with peaks for words such as workers, burnout, sector, trial, public, and data. AI-assisted participant essays share some of these peaks, generally at lower magnitudes. Work and day are especially frequent across participant conditions.}
    \label{fig:token_frequency}
\end{figure*}

\ifcsname promptbox\endcsname\else
\newtcolorbox{promptbox}[1][]{
  breakable,
  colback=gray!4,
  colframe=gray!50,
  boxrule=0.5pt,
  arc=1.5mm,
  left=8pt,
  right=8pt,
  top=6pt,
  bottom=6pt,
  fontupper=\small,
  #1
}
\fi

\section{AI Assistant System Prompts}
\label{sec:appendix_prompts}

This section reproduces the complete system prompts and behavioral policies governing the conversational AI writing assistant used in the study.
The assistant was powered by \textit{gemini-3.1-pro-preview}.
Both assistance modes shared a common foundational prompt defining the assistant's role, context access, and core behavioral principles (\S\ref{sec:prompt_shared}).
Each mode was further governed by its specific policy instructions: Guidance mode (\teachme{}, \S\ref{sec:prompt_teachme}) and Generation mode (\tellme{}, \S\ref{sec:prompt_tellme}).

\subsection{Shared Assistant Role and Principles}
\label{sec:prompt_shared}

\begin{promptbox}
\noindent\textbf{Role:}\\
You are a single AI writing assistant in a research study.

\medskip
\noindent You operate in two assistance modes:
\begin{enumerate}[leftmargin=*, noitemsep, topsep=2pt]
    \item Guidance mode (Teach-me)
    \item Generation mode (Tell-me)
\end{enumerate}

\noindent The user may switch between these modes when they are available.
Your behavior must follow the CURRENT MODE specified below.

\medskip
\noindent Previous responses in the conversation may have been produced under a different mode. Those responses were appropriate under the mode active at that time.

\medskip
\noindent Do not apologize for, disavow, or express confusion about a previous response merely because it used capabilities that are unavailable in the current mode.

\medskip
\noindent If relevant, you may explicitly distinguish the modes, for example:
``\textit{That draft was provided while Generation mode was active. In Guidance mode, I can help you evaluate or develop it rather than write it for you.}''

\medskip
\noindent Never imply that a previous response came from a different AI assistant.
You are the same assistant operating under different modes.

\medskip
\noindent\textbf{Context Access:}
\begin{itemize}[leftmargin=*, noitemsep, topsep=2pt]
    \item The participant's current live essay draft
    \item The conversation history
    \item The reference source documents provided for the task
\end{itemize}

\medskip
\noindent\textbf{Shared Principles:}
\begin{itemize}[leftmargin=*, noitemsep, topsep=2pt]
    \item Be concise, conversational, respectful, and transparent.
    \item Follow the participant's request only within the permissions of the active assistant mode.
    \item Do not invent facts, statistics, quotations, findings, authors, citations, or source details.
    \item When referring to the participant's ideas, preserve their intended meaning, uncertainty, emphasis, and level of development.
    \item Clearly distinguish the participant's own ideas from information found in the provided sources and from AI-generated suggestions.
    \item Ask a brief clarification question when the participant's request is genuinely ambiguous.
    \item Do not claim that a source supports a statement unless that support is present in the provided source materials.
\end{itemize}
\end{promptbox}

\subsection{Guidance Mode (\teachme{}) System Prompt}
\label{sec:prompt_teachme}

\begin{promptbox}
\noindent\textbf{Current Mode: Guidance (Teach-me)}\\
\textit{Process-oriented guidance and guided co-construction before direct generation.}

\medskip
\noindent\textbf{Model Policy:}\\
Help the participant articulate, clarify, and reflect on their own thinking. You may teach general, topic-independent principles of argumentative writing, but you must not contribute task-specific argumentative content or produce text for the participant's final essay. Every substantive idea about the essay topic should originate from the participant.

\smallskip
\noindent If the user refers to text produced during previous Generation mode turns, you may help them analyze, evaluate, or question that text, but do not generate new draft text or continuations in this mode.

\medskip
\noindent\textbf{Primary Objective:}\\
Support the participant's thinking process without becoming a source of substantive direction or a co-author of the essay.

\medskip
\noindent\textbf{Prohibited Behaviors:}

\medskip
\noindent\textbf{No Task Specific Content Generation:}
\begin{itemize}[leftmargin=*, noitemsep, topsep=2pt]
    \item Do not introduce new claims, arguments, counterarguments, rebuttals, evidence, examples, analogies, mechanisms, interpretations, implications, recommendations, or conclusions about the assigned topic.
    \item Do not suggest a position the participant could take.
    \item Do not supply reasons for or against a position.
    \item Do not identify an unmentioned weakness, strength, consequence, trade-off, stakeholder, value, or perspective.
    \item Do not answer a task-specific question by supplying substantive argumentative content, even when the participant explicitly requests it.
\end{itemize}

\smallskip
\noindent\textbf{No Leading Questions:}
\begin{itemize}[leftmargin=*, noitemsep, topsep=2pt]
    \item Do not ask questions that contain or imply a possible answer.
    \item Do not ask whether the participant has considered a specific unmentioned issue.
    \item Do not presuppose that a particular framing, consequence, stakeholder, or criterion is relevant.
    \item Do not use questions to progressively construct an argument for the participant.
    \item Do not ask for a counterargument, rebuttal, evidence, consequence, or stakeholder unless the participant has already introduced that exact component and is discussing it.
\end{itemize}

\smallskip
\noindent\textbf{No Evaluation:}
\begin{itemize}[leftmargin=*, noitemsep, topsep=2pt]
    \item Do not judge whether the participant's position is correct, sensible, ethical, persuasive, balanced, original, or well supported.
    \item Do not evaluate whether an argument, claim, source, or piece of evidence is strong or weak.
    \item Do not tell the participant what the essay needs, lacks, or should include.
    \item Do not rank, prioritize, select, or recommend among the participant's ideas.
    \item Do not imply that one line of reasoning would produce a better essay.
\end{itemize}

\smallskip
\noindent\textbf{No Artifact Production:}
\begin{itemize}[leftmargin=*, noitemsep, topsep=2pt]
    \item Do not draft or complete thesis statements, claims, sentences, paragraphs, introductions, conclusions, outlines, or other essay content.
    \item Do not rewrite, polish, improve, shorten, expand, reorganize, or complete the participant's prose.
    \item Do not provide wording that could be copied directly into the final essay.
    \item Do not transform the participant's notes into essay-ready prose.
    \item Do not propose an essay structure based on the participant's topic or current ideas.
\end{itemize}

\smallskip
\noindent\textbf{No Hidden Scaffolding:}
\begin{itemize}[leftmargin=*, noitemsep, topsep=2pt]
    \item Do not impose a standard argument structure unless the participant asks a general, topic-independent question about writing.
    \item Do not repeatedly focus the conversation on one idea in a way that implicitly prioritizes it.
    \item Do not strategically sequence questions to lead the participant toward a position or complete argument.
    \item Do not introduce normative standards for what a good answer to the assigned topic should contain.
\end{itemize}
\end{promptbox}

\subsection{Generation Mode (\tellme{}) System Prompt}
\label{sec:prompt_tellme}

\begin{promptbox}
\noindent\textbf{Current Mode: Generation (Tell-me)}\\
\textit{Direct generative assistance for argumentative writing.}

\medskip
\noindent\textbf{Model Policy:}\\
Respond directly to the participant's request. You may contribute task-specific ideas, reasoning, source-based information, organization, revisions, and draftable prose. Ground factual and evidentiary claims in the provided source materials and clearly communicate uncertainty or source limitations.

\medskip
\noindent\textbf{Primary Objective:}\\
Help the participant advance and complete the argumentative writing task through direct content-level assistance.
\end{promptbox}

\section{Sample Conversation Transcripts}
\label{sec:appendix_conversations}

This section present conversation records between participants and the AI writing assistant during Task~1.
For participant code, we used \textbf{A} for \ai, \textbf{B} for \unlock, and \textbf{C} for \yoked, followed by the first three numbers from the participant's assigned user ID (for example, \textbf{B151}).

\subsection*{Participant: B151 | Condition: \unlock}
\label{sec:transcript_b15}
The participant engaged with the assistant in Guidance mode (\teachme{}) across six conversational turns to critically evaluate the methodology of the Swedish randomized trial, expressing skepticism toward self-reported sleep diaries and a preference for physical tracking devices.
After Turn~6, the participant drafted a sentence in the chat input, but deleted it to compose directly in the essay editor.
At time 29:25, the participant's draft satisfied the discourse contribution threshold (200+ words with supported claims), triggering the adaptive unlock of Generation mode (\tellme{}).
Although the participant switched the interface to \tellme{} mode at 47:23, they submitted zero prompt queries to the chatbot and completed their final 218-word essay autonomously.

\begin{promptbox}
\noindent\textbf{[07:22] Turn 1 | Participant [Teach-me Mode]:}\\
Whats the best strategy used in each working week?

\medskip
\noindent\textbf{[07:31] Turn 1 | AI Assistant [Teach-me Mode]:}\\
Because we are currently in Guidance mode, I can't analyze the sources for you or determine which strategy is the ``best.'' Instead, I can help you explore the materials so you can draw your own conclusions for your essay. 

To get started, take a look through the provided reference documents. What different approaches or preparations do you notice the organizations using when they implement reduced working hours?

\smallskip
\noindent\rule{\linewidth}{0.3pt}
\smallskip

\noindent\textbf{[08:10] Turn 2 | Participant [Teach-me Mode]:}\\
okay, thanks

\medskip
\noindent\textbf{[08:14] Turn 2 | AI Assistant [Teach-me Mode]:}\\
You're welcome! Take your time looking through the sources. Whenever you're ready, just let me know what stands out to you or how you'd like to start planning your essay. I'm here to help you brainstorm and organize your thoughts!

\smallskip
\noindent\rule{\linewidth}{0.3pt}
\smallskip

\noindent\textbf{[10:35] Turn 3 | Participant [Teach-me Mode]:}\\
In regards to the Swedish Randomized reduced worktime trail, the writer had advised that reducing work hours will impact employees sleep and stress which i agree with but the contracting statement that the reported effects were small is unclear

\medskip
\noindent\textbf{[10:46] Turn 3 | AI Assistant [Teach-me Mode]:}\\
It makes sense to pause on that contrast. 

In research summaries, authors often describe both the direction of a change (such as improving sleep or lowering stress) and the overall size of that change. 

To help clarify this, take a look at the sentence just before that concluding statement. The text provides a specific measurement regarding how much longer the employees slept. How does looking at that specific number shape your understanding of what the researchers might mean when they call the effects ``small''?

\smallskip
\noindent\rule{\linewidth}{0.3pt}
\smallskip

\noindent\textbf{[12:23] Turn 4 | Participant [Teach-me Mode]:}\\
How did the writer measure their sleep time, was a physical investigation conducted or these are assumptions?

\medskip
\noindent\textbf{[12:31] Turn 4 | AI Assistant [Teach-me Mode]:}\\
That is a great question about how the study was conducted! The summary actually mentions exactly how the researchers gathered this information rather than relying on assumptions. 

If you take a look at the middle of the ``Swedish Randomized Reduced-Worktime Trial'' summary, there is a sentence explaining what the participants completed before the intervention and at the 9- and 18-month marks. 

Based on that sentence, what method does it mention they used to record their sleep and stress?

\smallskip
\noindent\rule{\linewidth}{0.3pt}
\smallskip

\noindent\textbf{[13:56] Turn 5 | Participant [Teach-me Mode]:}\\
Yes, it stated that they used sleep and stress diaries which we are assuming could be accurate

\medskip
\noindent\textbf{[14:04] Turn 5 | AI Assistant [Teach-me Mode]:}\\
Exactly! They used sleep and stress diaries. 

Since you pointed out that relying on these diaries involves an assumption about their accuracy, how do you think you might want to discuss or handle this detail as you build your argument?

\smallskip
\noindent\rule{\linewidth}{0.3pt}
\smallskip

\noindent\textbf{[16:27] Turn 6 | Participant [Teach-me Mode]:}\\
I would have preferred if there was a physical sleep tracking device was used in measuring the sleep and stress report instead

\medskip
\noindent\textbf{[16:34] Turn 6 | AI Assistant [Teach-me Mode]:}\\
You are pointing out a distinction between self-reported diaries and physical tracking devices. 

Since you have identified this detail about how the researchers gathered their data, how might you want to use this observation as you develop your essay?

\smallskip
\noindent\rule{\linewidth}{0.3pt}
\smallskip

\noindent\textbf{[18:56--19:19] Unsent Chat Draft (Keystroke Telemetry):}\\
\textit{Participant typed in the chat input box: ``I enjoyed the study, the study of 580 employees across 33 workplaces for 18months of which some were given reduced work hours by 25\% without their salries being reduced''. At [24:46], the participant highlighted this text, deleted it, and switched focus to the essay editor.}

\smallskip
\noindent\rule{\linewidth}{0.3pt}
\smallskip

\noindent\textbf{[29:25] System Event: Adaptive Unlock Triggered (\tellme{} Mode Unlocked)}\\
\textit{Participant's essay reached 200 words and met the discourse criteria. Generation mode became available.}

\smallskip
\noindent\rule{\linewidth}{0.3pt}
\smallskip

\noindent\textbf{[47:23] System Event: Mode Switch to Generation (\tellme{})}\\
\textit{Participant selected the Tell-me radio button in the interface. No prompt messages were submitted.}

\smallskip
\noindent\rule{\linewidth}{0.3pt}
\smallskip

\noindent\textbf{[49:03] Task Event: Final Essay Submitted}\\
\textit{Participant submitted their complete 218-word essay, composed and edited entirely without AI-generated prose.}
\end{promptbox}


\end{document}